\documentclass[tradiabstract]{aa} 
\usepackage{graphicx}
\usepackage{txfonts}
\usepackage{natbib}
\newcommand\uchii{\mbox{UCH\scriptsize{II}\normalsize}}

\newcommand\msol{\mbox{M$_\odot$}}

\newcommand\cmc{\mbox{cm$^{-3}$}}

\newcommand\asec{\mbox{$^{\prime\prime}$}}%
\newcommand\adeg{\mbox{$^\circ$}}%
\begin{document}
   \title{Fragmentation and mass segregation in the massive dense cores of Cygnus X}
   \titlerunning{Fragmentation in the massive dense cores of Cygnus X}


   \author{S. Bontemps,
          \inst{1}\fnmsep\inst{2}\fnmsep\inst{3}
	  F. Motte,\inst{3}
          T. Csengeri,\inst{3}
	  N. Schneider\inst{3}
          }

   \institute{CNRS/INSU, Laboratoire d'Astrophysique de Bordeaux, UMR 5804, BP 89, 33271 Floirac cedex, France\\
              \email{bontemps@obs.u-bordeaux1.fr}
         \and
             Universit\'e de Bordeaux, Observatoire Aquitain des Sciences de l'Univers, Bordeaux, France
         \and
             Laboratoire AIM, CEA/DSM, IRFU/Service d'Astrophysique, 91191 Gif-sur-Yvette Cedex, France
                  }

   \date{Received ; accepted }


  \abstract
  {Massive dense cores (MDCs) are the high-mass equivalent of the so-called dense cores in nearby star-forming regions. With typical sizes of 0.1~pc, they could be either  single high-mass protostars, or proto-clusters on the verge to form a cluster of low-mass stars. Here we present high angular resolution continuum observations obtained with the IRAM Plateau de Bure interferometer at 1.3 and 3.5~mm towards the six most massive and young (IR-quiet) dense cores in the Cygnus X complex. Located at only 1.7 kpc, the Cygnus X region offers the opportunity of reaching small enough scales (of the order of 1700 AU at 1.3~mm) to separate individual collapsing objects, and thus to observe and constrain the result of the fragmentation process. The cores are sub-fragmented with a total of 23 fragments inside 5 cores. This is an indication that the monolithic collapse view may not apply to these cores. However, the most compact MDC, CygX-N63, stands out as a single object, suggesting it could actually be a single massive protostar with an envelope mass as large as $\sim60\,$\msol. The fragments in the other cores have sizes and separations similar to low-mass pre-stellar and proto-stellar condensations in nearby protoclusters, and are most probably self-gravitating objects (M $>$ M$_{\rm vir}$)  and are either pre-stellar condensations or Class~0 young stellar objects.  A total of 8 out of these 22 protostellar objects are found to be probable precursors of OB stars in addition to CygX-N63, with their envelope masses ranging from 6 to 23$\,$\msol$\,$ inside a FWHM of 3000 AU. The level of fragmentation is therefore globally higher than in the turbulence regulated, monolithic collapse scenario, but it is also not as high as expected in a pure gravo-turbulent scenario where the distribution of mass is dominated by low-mass protostars/stars. Here, the fractions of the total MDC masses in the high-mass fragments are reaching values as high as 28, 44, and 100~\% in CygX-N12, CygX-N53, and CygX-N63, respectively, much higher than what an IMF-like mass distribution would predict. The increase of the fragmentation efficiency as a function of average density in the MDCs is proposed to be due to the increasing importance of self-gravity leading to gravitational collapse at the scale of the dense cores. At the same time, the observed MDCs tend to fragment into a few massive protostars within their central regions.  We are therefore probably witnessing here the primordial mass segregation of clusters in formation. The physical origin of the fragmentation into a few high-mass objects is not clear yet, and will be further investigated in the future by investigating the kinematics of the protostars, and by measuring the importance of magnetic support, to check the two possible scenarios: early competitive accretion or magnetic supression of fragmentation during the global collapse.}


   \maketitle
%

\section{Introduction}
\label{intro}

The detailed physical processes at the origin of high-mass stars (here defined as ionizing
OB stars; i.e. earlier than B3 or M$_\star >\,8$\,M$_\odot$) are still 
to be fully recognized and understood. 
Whether they form through a scaled-up version of the
low-mass star formation process or require a specific process
is still an open issue. 
The formation process might be particular for two main reasons. First, the radiation pressure of the ionizing protostar could stop accretion when it reaches a mass of  $\sim 10\,$M$_\odot$. With a moderate accretion rate ($\sim 10^{-5}\,$M$_\odot$/yr), the star could at best reach $\sim 30\,$M$_\odot$ \citep[see][and references therein]{yorke&sonnhalter2002}.
Second, OB stars
are found to form in clusters and their formation could originate from collective effects. Competitive accretion is one of the scenarios which require cluster forming conditions since the accretion efficiencies and rates depend on the  
gravitational potential of the proto-cluster  \citep{bonnell&bate2006}. 
The radiation pressure may not be a problem after all if the accretion rates are in practice much higher than usually measured in low-mass protostars (typically between 10$^{-6}$  and 10$^{-5}\,$M$_\odot$/yr). 
High infall rates could actually be expected in the classical theory of star formation applied to low-mass stars if the initial conditions for collapse significantly depart from
a singular $r^{-2}$ sphere or if the collapsing conditions are reached in a dynamical way (\citealp[see][]{larson1969}; \citealp{whitworth&summers1985}; 
\citealp{foster&chevalier1993}; \citealp{henriksen1997}; \citealp{chabrier2003}).
Alternatively, \citet{mckee&tan2002} propose that even in a slow, quasi-static evolution of star-forming clumps, large accretion rates
could be due to the high level of turbulence. 
However, a continuous injection of turbulence in these high density regions would then be required. Turbulence is also usually believed to be at the origin of the hierarchical structure of molecular clouds \citep[e.g.][]{falgarone2004} and may therefore regulate the fragmentation of clouds from large scales down to collapsing cores   (\citealp[see][]{padoan&nordlund2002}; \citealp{chabrier2003}).  The turbulent fragmentation can adequately reproduce the observed core mass functions which mimic the IMF of stars \citep{klessen2001}. In this description, the most massive stars would originate from the rare high-mass fragments which could decouple from their turbulent environment. This however implies that massive stars would be statistically dispersed with large distances between them, in contradiction with their observed tendency to cluster. In a purely gravitational view, the typical stellar masses are dictated by the Jeans masses which are usually of the order of a fraction of a solar mass. This can explain the average stellar masses, but has problems with explaining the highest masses. The latter are often two orders of magnitude  larger than the Jeans masses. Some additional effects are therefore needed to fully understand how fragmentation proceeds to produce massive cores, or to enable specific conditions to grow massive stars from low-mass collapsing fragments.

The final fragmentation leading to individual collapsing protostars inside the dense massive clumps is therefore the central question for understanding the origin of massive stars. In the slow evolution of turbulent clouds promoted by \citet{mckee&tan2002} the level of fragmentation is reduced and should lead to stable (relatively long living) massive cores which would collapse into a single star. In the dynamical view, the clumps are expected to fragment into a large number of low-mass fragments  \citep[e.g.][]{bonnell2007} which may grow in mass in a runaway process, in the center of the forming cluster when the gravitational dwell is dominated by the protostellar and stellar components. The fragmentation inside the most massive dense cores at the scale of a few 0.1 pc, i.e. the best candidates to be proto-clusters in their earliest phases, is highly discriminating for the theoretical schemes in competition. 

Since massive stars are rare, and the lifetime of their earliest phases is short, the protostellar precursors of massive stars are elusive objects. Large scale and systematic surveys are mandatory to find them. The early works obtained in the vicinity of luminous IRAS or maser sources in the whole galactic plane (\citealp{molinari1996}, \citealp{plume1997}, \citealp{beuther2002a}, \citealp{mueller2002}, \citealp{faundez2004}), and more recently inside the dark filaments seen in the infrared surveys of the Galactic plane, the so-called IRDCs \citep[e.g.][ and references therein]{simon2006}, have found a number of candidates to host such precursors (\citealp{rathborne2006}, \citealp{pillai2006}). These objects are however spread over the whole Galaxy, at distances usually larger than 3 kpc, and their detection are not based on complete imaging of star-forming complexes, not providing a representative view of these not well studied cold phases. In Cygnus X, \cite{motte2007} could achieve such a complete survey of a molecular complex rich enough to provide a sample of massive dense cores (hereafter MDCs). Cygnus X is a nearby massive complex with a distance of only 1.7 kpc (\citealp{schneider2006}, \citealp{schneider2007}).
A total of 40 MDCs has been discovered. These cores are dense, and massive on a scale of $\sim\,0.1\,$pc and can thus be expected to be representative of the earliest phases of the formation of high-mass stars. 
Even the MDCs which are not seen as dark silhouettes on the IR background (IRDCs) are recognized.
Following the nomenclature by \citet{williams2000}, we note that, in terms of sizes, the $\sim\,0.1\,$pc MDCs of Cygnus X should be considered as cores. We here adopt the older and even more classical term of ``dense cores" for the $\sim\,0.1\,$pc entities even if they might be sub-fragmented and may form a small cluster. The term ``fragments" is then used for smaller structures at the scale of a few 0.01$\,$pc inside these dense cores, and which could correspond to individual prestellar or protostellar envelopes such as the ones found in the nearby and low-mass clusters in formation (e.g. \citealp{motte1998}).

We present the first results of a systematic study at high spatial resolution of the MDCs of Cygnus X. The 1 and 3~mm continuum maps of 6 IR-quiet MDCs obtained with the Plateau de Bure interferometer are discussed here. The accompanying line emission obtained in the same time will be included in forthcoming papers dedicated to outflows for the CO and SiO lines, and to the dense molecular gas inside the MDCs for the H$^{13}$CO$^+$ and H$^{13}$CN lines. The source selection and the observations are explained in  Sect.~\ref{obs}.  The continuum maps as well as pointed observations in N$_2$H$^+$ towards each MDCs obtained with the IRAM 30$\,$m telescope are shown in Sect.~\ref{results}. In Sect.~\ref{discussion}, we discuss the obtained properties of the fragmentation of these MDCs in the framework of understanding the origin of high-mass stars.



\section{Source selection and observations}
\label{obs}

\subsection{IR-quiet MDCs in Cygnus X}

From the complete imaging survey of the high column density regions of Cygnus X in the 1.3$\,$mm continuum with MAMBO on the IRAM 30m telescope, 
\citet{motte2007} recognized 129 dense cores with masses ranging from 4 to 950~M$_\odot$. Among these, a total of 40 were found to be more massive than 40~\msol ~and were identified to be probable precursors of high-mass stars in Cygnus X.  With an average size of 0.13$\,$pc,
and an average density of $1.9\times10^5\,$\cmc, these MDCs have sizes only slightly larger
than low-mass dense cores in nearby molecular clouds  (\citealp{ward-thompson1994},  \citealp{bacmann2000}, \citealp{evans2001}, \citealp{ward-thompson2007}). But they are on average almost 10 times more dense and 20 times more massive than their low-mass counterparts (see Tab.~4 in \citealp{motte2007}).
Despite such high densities and masses, which should lead to intense 
star formation activity, as many as 17 out of 40 cores do not emit or emit only weakly in the infrared (IR-quiet). This lack of strong infrared emission could indicate that these regions were not yet forming massive stars. On the other hand, strong SiO emission is systematically detected towards them. The intensities and profiles of these
SiO lines indicate the presence of powerful outflows most probably driven by high-mass protostars. These properties are very similar to the properties of low-mass Class~0 young stellar objects (hereafter YSOs) as dense, cold objects driving powerful outflows (e.g. \citealp{awb1993}). A critical comparison suggests that the IR-quiet MDCs are either a scaled-up version (in size and mass) of low to intermediate mass Class~0 YSOs or correspond to clusters with a significant number of low-mass Class~0 YSOs. 
These clustered Class~0s could further compete for mass in the center of the cluster if their crossing time is much shorter than their collapsing time. Alternatively the MDCs may be sub-fragmented into a fewer number of high-mass Class~0 YSOs to be identified at smaller scales. In all scenarios, these IR-quiet MDCs constitute a sample of prime targets for investigating the earliest phases of massive star and cluster formation. 

\subsection{A representative sample of 6 IR-quiet MDCs}

\begin{figure}
   \centering
   \includegraphics[width=6.5cm,clip,angle=0]{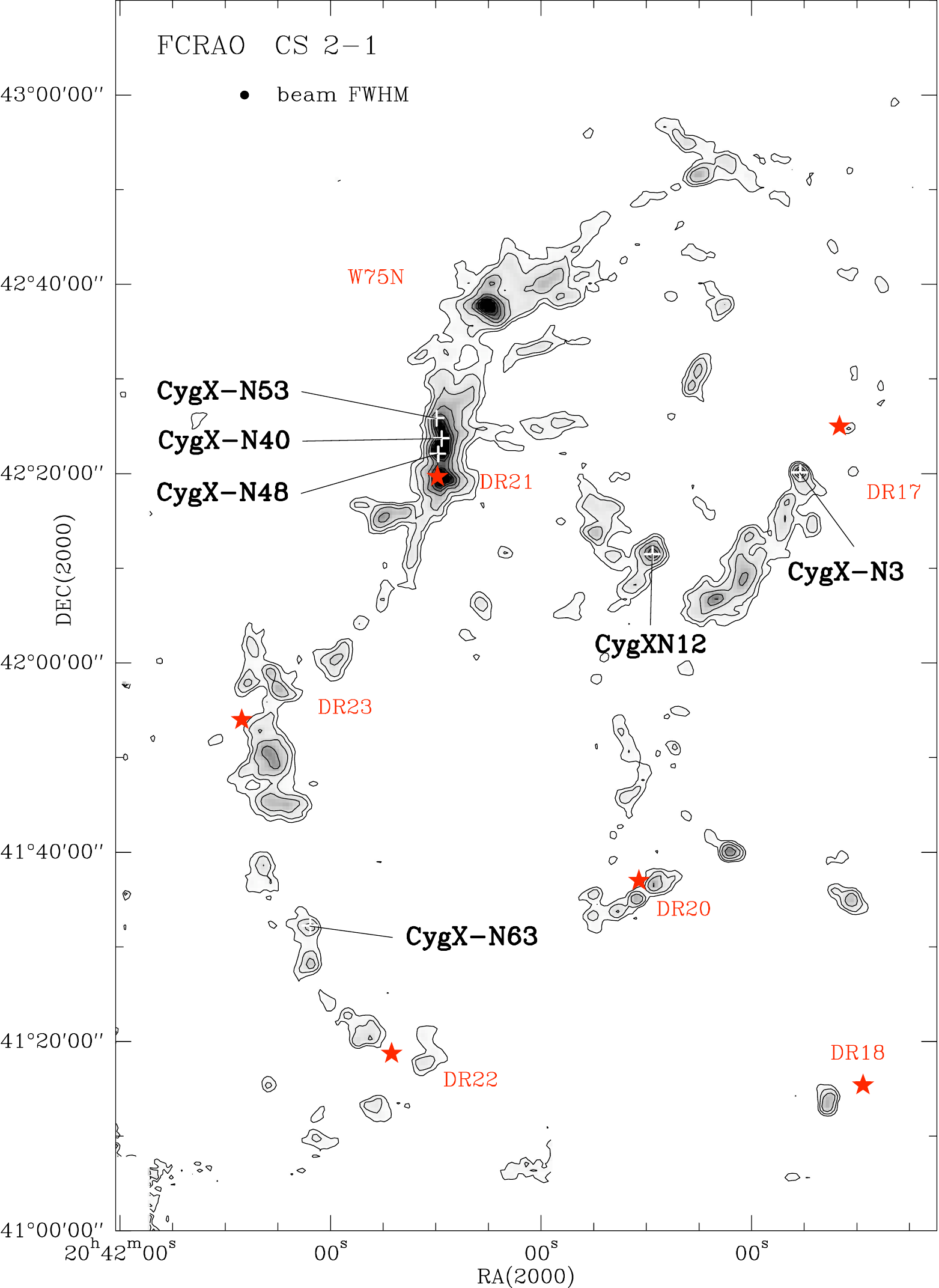}
   \caption{Overview of the northern part of Cygnus X indicating the location of the 6 MDCs from \citet{motte2007}. The contour map displays the integrated emission in CS($2-1$) as observed with the FCRAO (Schneider et al. 2009, in prep). Note that 3 of the 6 MDCs (CygX-N40, CygX-N48, and CygX-N53) are clustered inside the most massive region of Cygnus X, the DR21 filament. The red stars indicate the centers of the HII regions DR17 to DR23 from \citet{downes&rinehart1966}.}
              \label{overview}%
   \end{figure}

We have selected the 6 most massive IR-quiet MDCs of Cygnus X. The names and coordinates from \citet{motte2007} are listed in Tab.~\ref{obs_summary}. An overview of the locations of these MDCs is given in Fig.~\ref{overview}. Three of the cores (CygX-N40, CygX-N48, and CygX-N53) are located in the DR21 filament which is the most massive and dense region of Cygnus X, and a well-known region of massive star formation. CygX-N40,  and CygX-N48 correspond to the submillimeter sources DR21(OH)-N2, DR21(OH)-S, while CygX-N53 is situated close to FIR3 \citep{chandler1993}. The other 3 selected MDCs have been discovered by the MAMBO survey and are situated in more isolated, but still prominent molecular clumps bright in CS emission (Fig.~\ref{overview}). CygX-N3 in the western part, is located close to DR17 which is an H{\small II} region excited by two OB clusters (clusters \# 12 and 14 in \citealp{leduigou2002}). These clusters are shaping the cloud and  CygX-N3 corresponds to the tip of a pillar-like cloud.  CygX-N12 is also situated in a cometary shape cloud probably influenced by the same OB clusters but the cloud seems to be less compressed from outside than CygX-N3. Finally CygX-N63 is placed in the south of DR21 in the DR22-DR23 filament \citep[see][]{schneider2006}. It can be considered as the most isolated MDC of the sample.  
\begin{table*}
\begin{minipage}[t]{\columnwidth}
\caption{Observing parameters and obtained beam and rms.}
\label{obs_summary}
\centering
\renewcommand{\footnoterule}{}  
\begin{tabular}{ccccccc}
\hline \hline
       &    \multicolumn{2}{c}{Phase center}  & \multicolumn{2}{c}{Synthesized beam} & \multicolumn{2}{c}{Obtained rms}            \\
Source        &    \multicolumn{2}{c}{(J2000)}       & \multicolumn{2}{c}{[arcsec]}         & \multicolumn{2}{c}{[{\rm mJy/beam}]}    \\
name & R.A.     &  Decl.                    &     1.3 mm     &            3.5 mm  &     1.3 mm     &            3.5 mm    \\
\hline
 CygX-N3   & 20 35 34.1   &  42 20 05.0      &  1.09$\,\times\,$0.92  &  2.66$\,\times\,$2.20   &     1.18               &  0.212  \\
 CygX-N12  & 20 36 57.4   &  42 11 27.5      &  1.14$\,\times\,$0.83  &  3.02$\,\times\,$2.28   &     1.95               &  0.221  \\
 CygX-N40  & 20 38 59.8   &  42 23 42.0      &  1.08$\,\times\,$0.92  &  2.62$\,\times\,$2.17   &     1.04               &  0.171  \\
 CygX-N48  & 20 39 01.5   &  42 22 04.0      &  1.12$\,\times\,$0.91  &  2.81$\,\times\,$2.33   &     2.25               &  0.340  \\
 CygX-N53  & 20 39 03.1   &  42 25 50.0      &  1.05$\,\times\,$0.87  &  3.16$\,\times\,$2.57   &     1.94               &  0.246  \\
 CygX-N63  & 20 40 05.2   &  41 32 12.0      &  1.15$\,\times\,$0.83  &  3.05$\,\times\,$2.24   &     4.25               &  0.332  \\
\hline
\end{tabular}
\end{minipage}
\end{table*}

\subsection{Interferometric observations of the 1  and 3~mm continuum}

We used the IRAM\footnote{IRAM is supported by INSU/CNRS (France), MPG (Germany) and IGN (Spain).} Plateau de Bure Interferometer (hereafter PdBI) to image at high resolution the dust continuum simultaneously at 3.5 and 1.3~mm. The 1~mm and 3~mm receivers were tuned to 230.54 and 86.61 GHz respectively with two 320 MHz units placed at 230.41 and 230.67 GHz and 86.48 and 86.74 GHz respectively. In addition to the continuum, the emission in 4 molecular lines has also been correlated: $^{12}$CO($2-1$), SiO($2-1$), H$^{13}$CO$^+$($1-0$), and H$^{13}$CN($1-0$) at 230.54, 86.85, 86.75, and 86.34 GHz respectively.

The observations were done in track-sharing mode with two targets per track with the following pairs:
CygX-N48/CygX-N53, CygX-N3/CygX-N40, and CygX-N12/CygX-N63. The D configuration tracks were obtained between June and October 2004 with 5 antennas with baselines ranging from 24 m to 82 m. The C configuration tracks were obtained in November and December 2004 with 6 antennas in 6Cp with baselines ranging from 48 m to 229 m. We mostly used as phase calibrator the bright nearby quasar 2013+370 and as flux calibrator the evolved star MWC349 which is located in Cygnus X.

In order to favor the spatial resolution, the maps were cleaned using the uniform weighting. The resulting synthesized beam and rms in the continuum are summarised in Tab.~\ref{obs_summary} together with the field names and centers of phase. The cleaning components were searched in the whole area of the primary beams. No polygon was used in order to avoid introducing any bias in the resulting maps of emission.

\subsection{IRAM 30~m observations of N$_2$H$^+$ 1$\to$0}

The N$_2$H$^+$ 1$\to$0 line at 93.176265 GHz was observed in september
2003 using the B100 receiver and the VESPA correlator at the IRAM 30m
telescope. The average system temperature was 182 K and the average
rms of the spectra is 0.19 K.  The data are corrected for the main
beam efficiency of 0.78 and have a velocity resolution of 0.13 km
s$^{-1}$, and a beamsize of 29\asec.

   \begin{figure*}
    \centering
    \includegraphics[width=6.5cm,clip,angle=0]{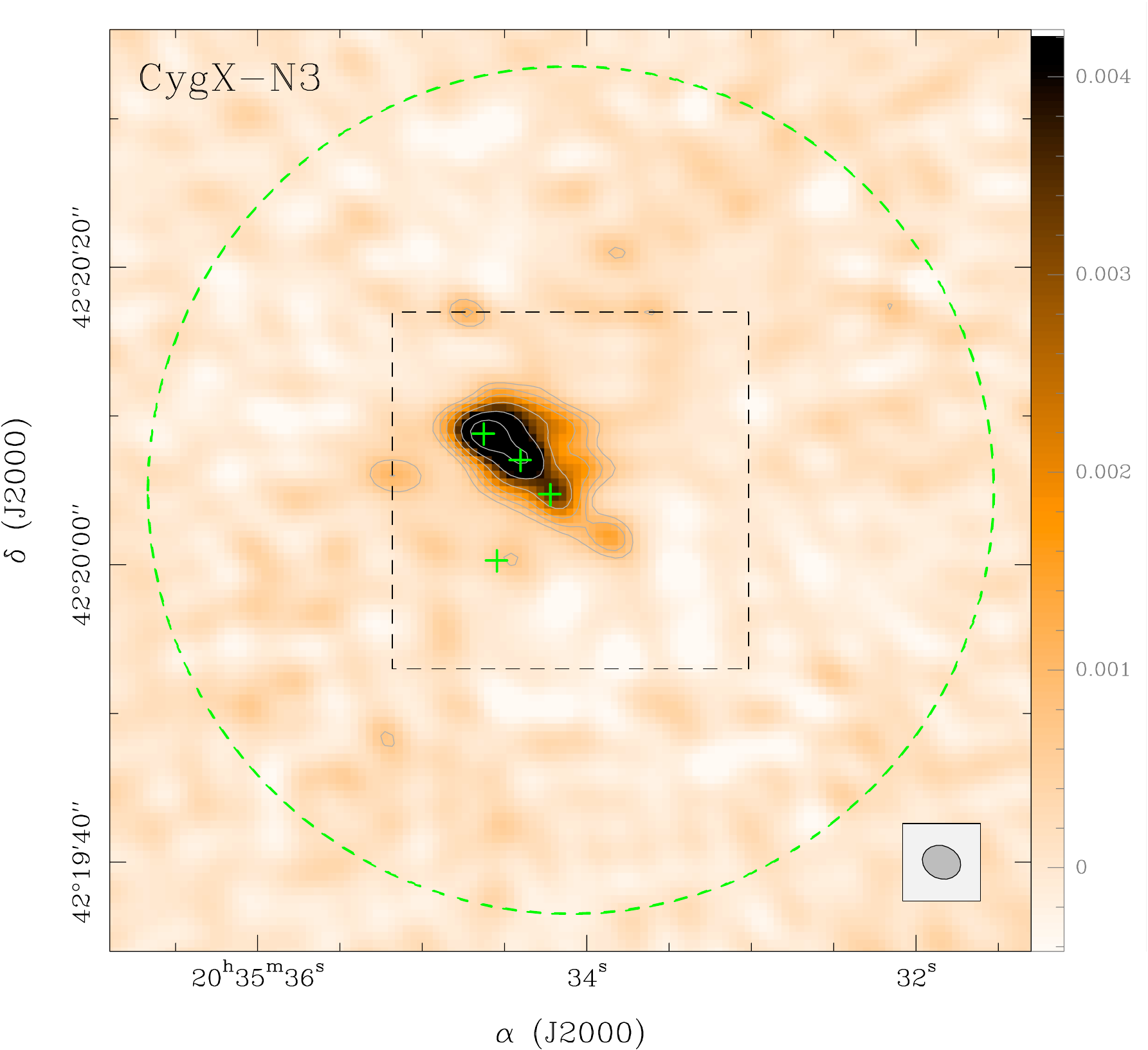}\hspace{0.23cm} 
    \includegraphics[width=6.5cm,clip,angle=0]{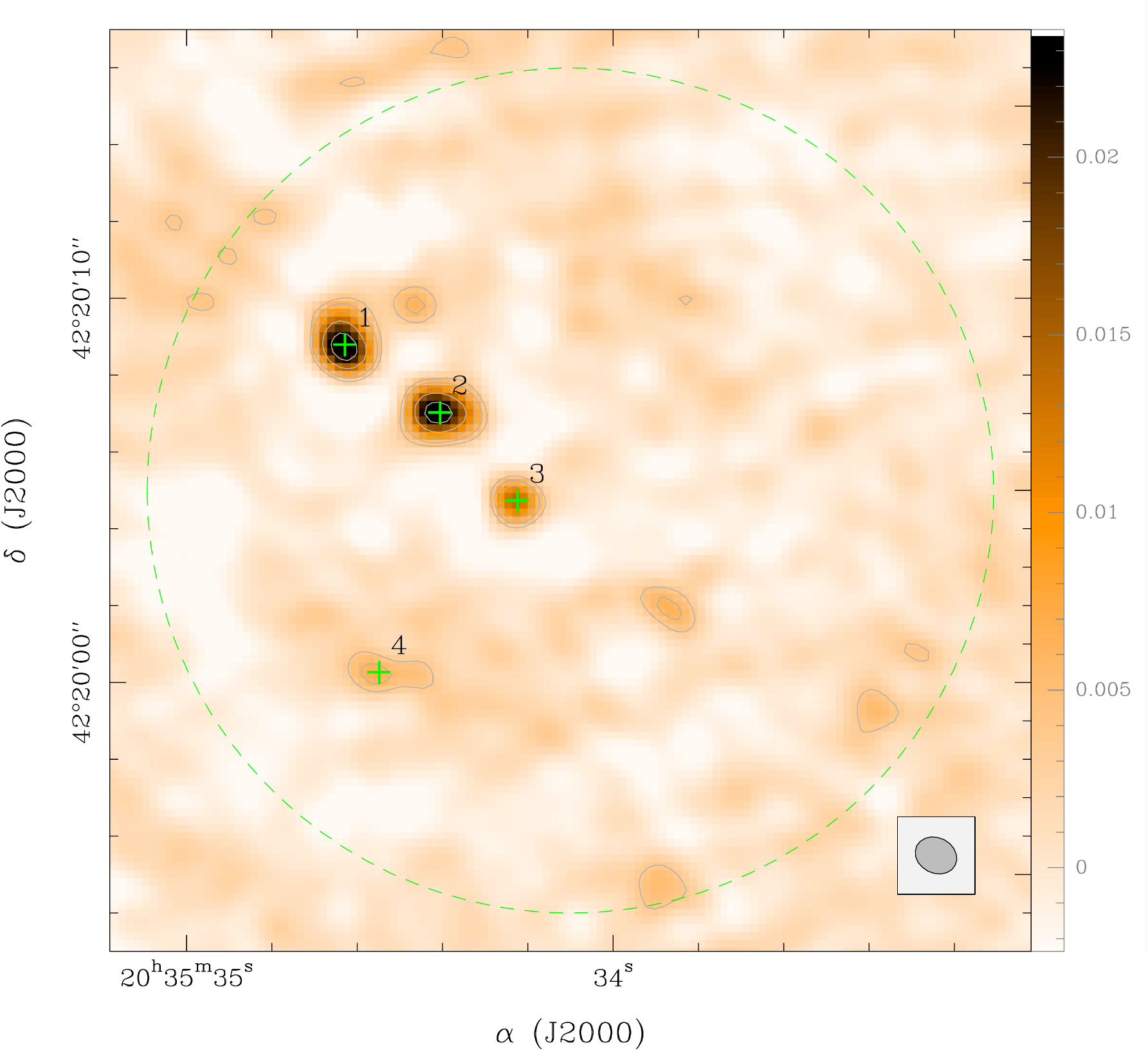}\\
    \includegraphics[width=6.5cm,clip,angle=0]{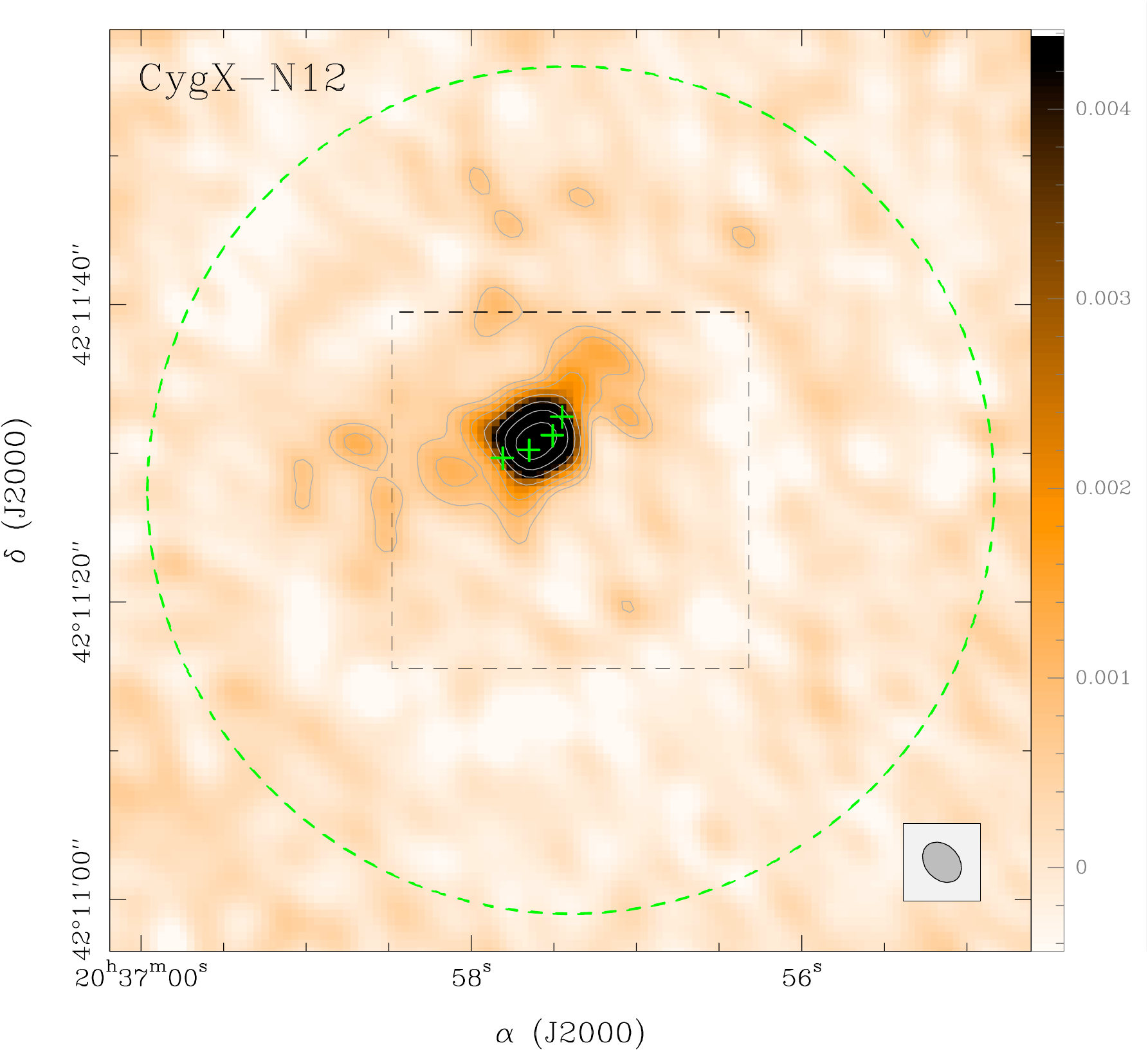}\hspace{0.23cm} 
    \includegraphics[width=6.5cm,clip,angle=0]{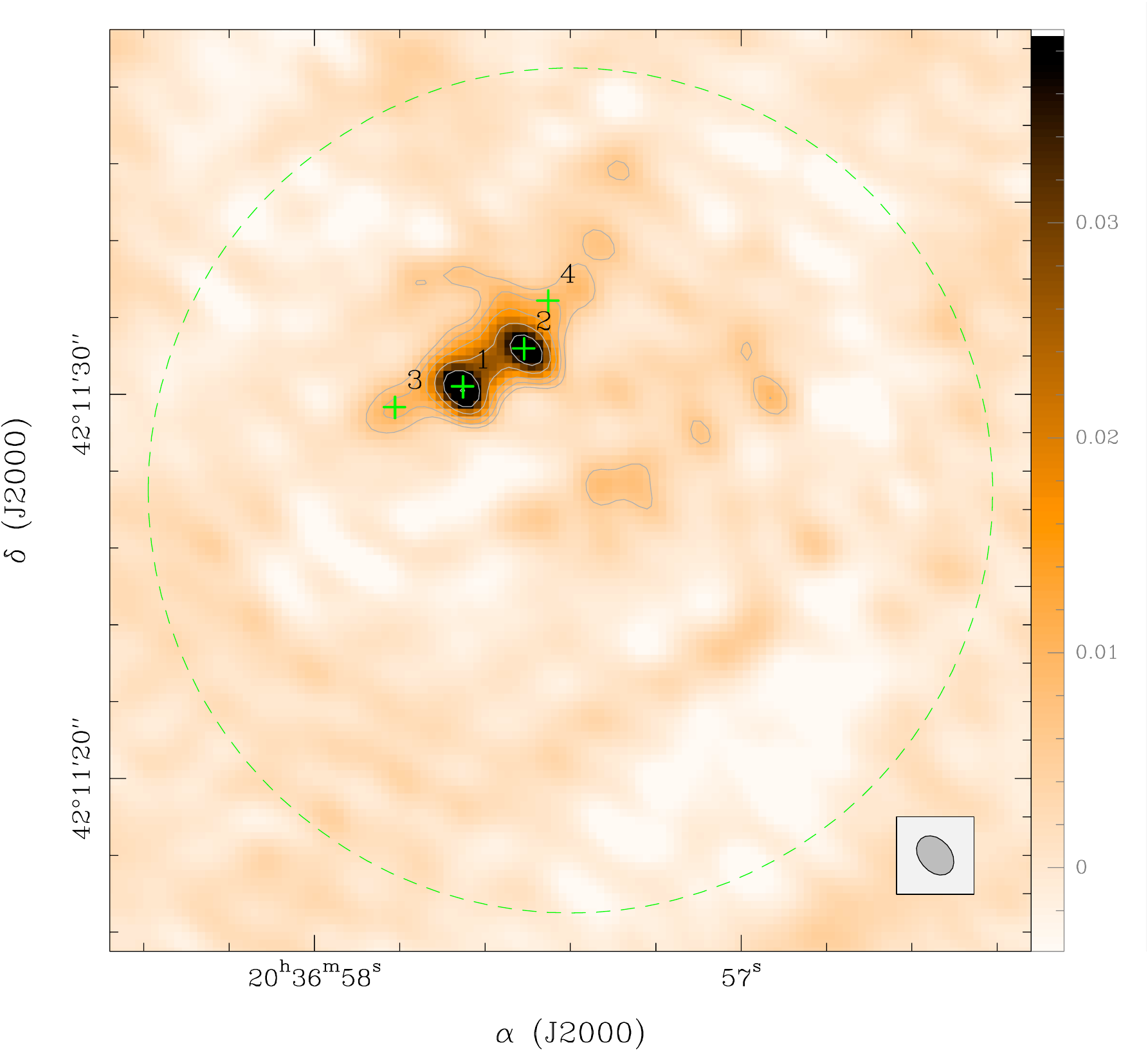}\\
    \includegraphics[width=6.5cm,clip,angle=0]{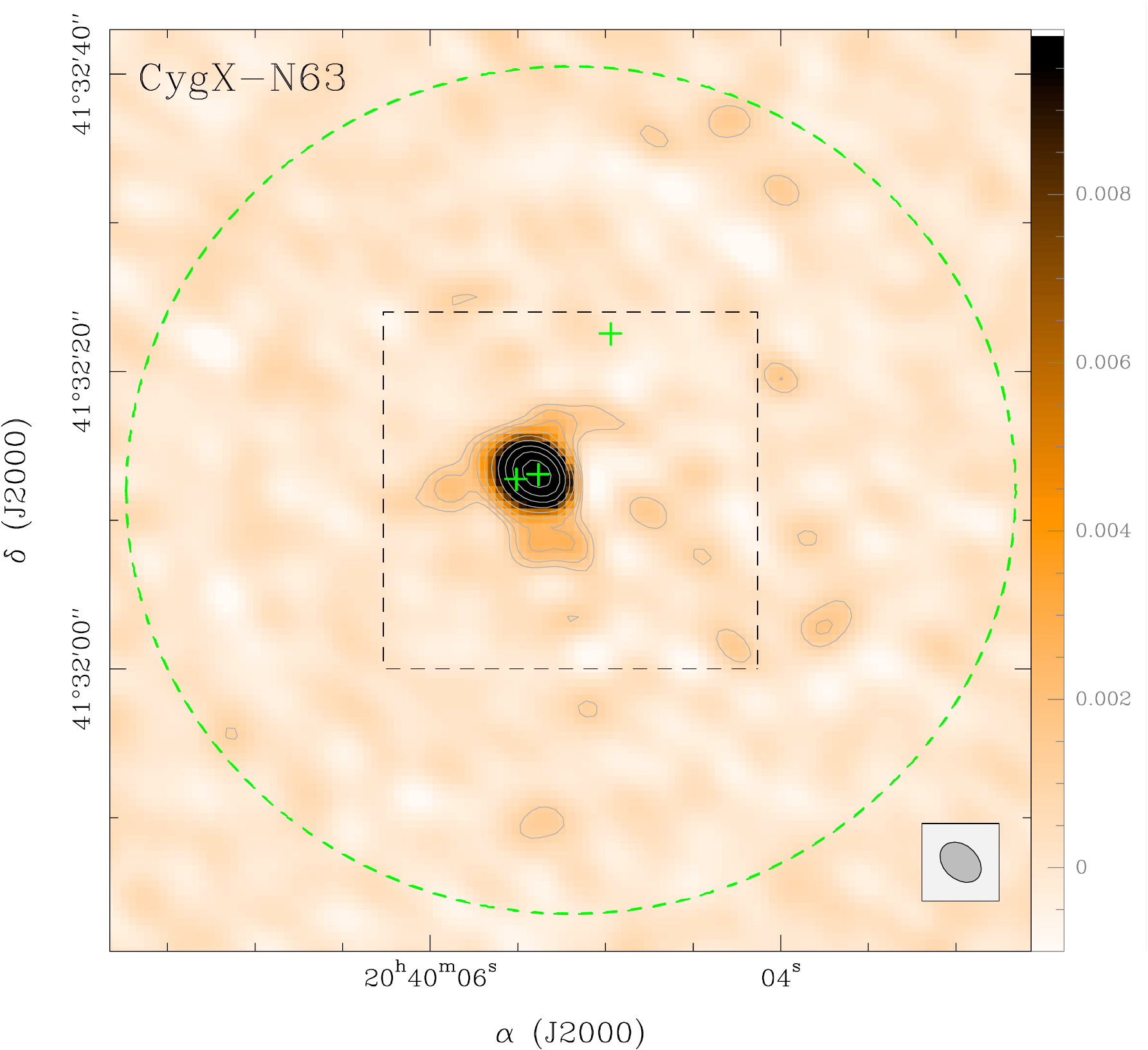}\hspace{0.23cm} 
    \includegraphics[width=6.5cm,clip,angle=0]{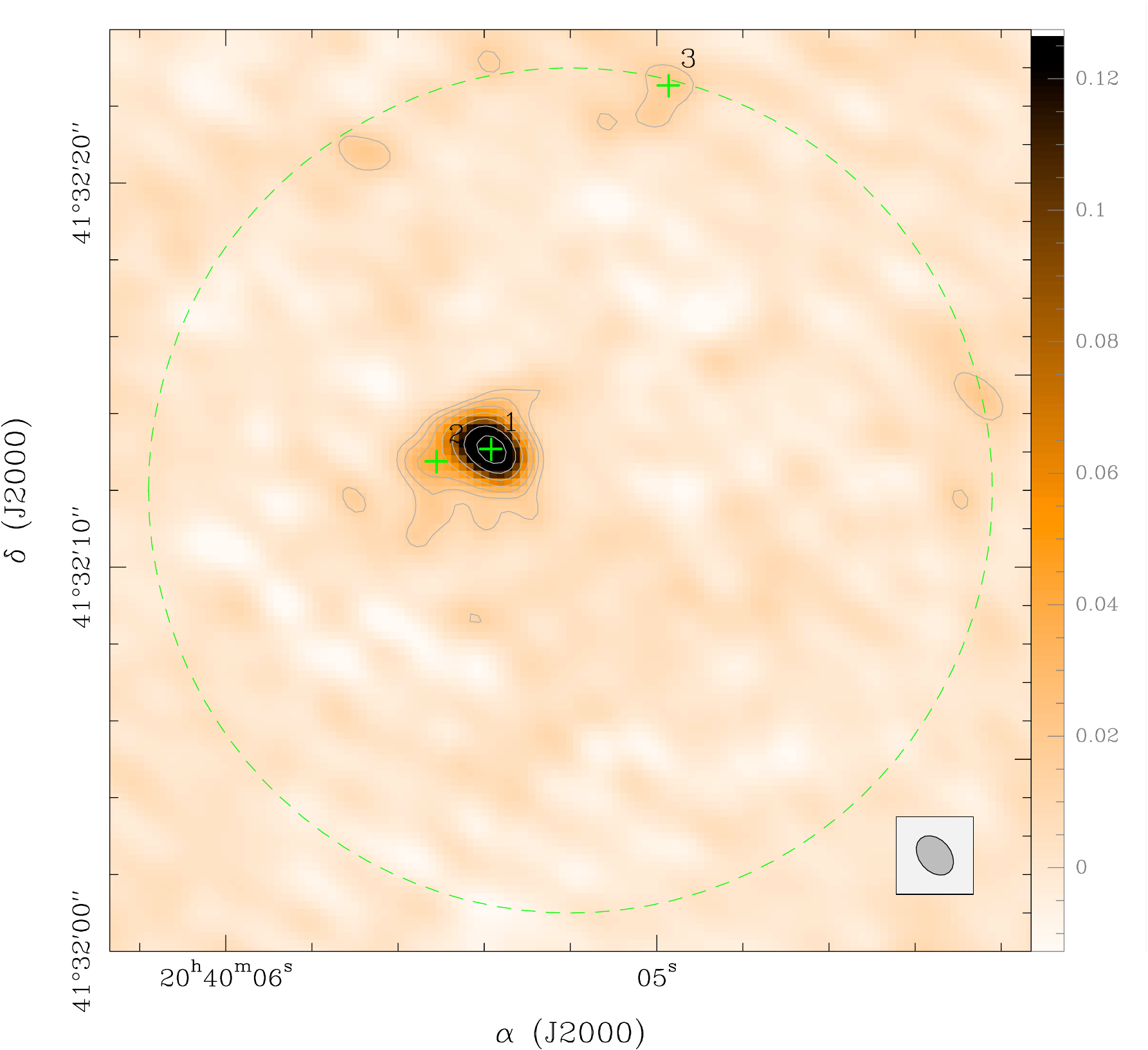}
    \caption{Maps of the continuum at 3.5$\,$mm (left) and 1.3$\,$mm (right) obtained with the PdBI for the three more isolated MDCs (see text) . The color images have linear scales from $-2\,\sigma$ to $+20\,\sigma$  for the 4 top panels, and from $-3\,\sigma$ to $+30\,\sigma$ for the 2 lower ones where $\sigma$ is the resulting rms in each field (see Tab.~\ref{obs_summary})(the unit on the right of the plots is Jy/beam). The (logarithmic) contour levels are 3$\,\sigma$, 4.8$\,\sigma$, 7.5$\,\sigma$, 12$\,\sigma$, 19$\,\sigma$, 30$\,\sigma$, 75$\,\sigma$, and 120$\,\sigma$. The crosses and numbers indicate the locations and the names of the fragments at  1$\,$mm (see Tab.~\ref{list_sources}). The synthetised beams are displayed in the bottom right corners. The dashed circles indicate the primary beams. In the  3.5$\,$mm maps (left), the name of the MDC is given, and the dashed squares indicate the areas covered  at 1.3$\,$mm (right). Note the similar behavior of the 3 MDCs. A compact core is observed at 3.5~mm which splits into smaller fragments at 1.3~mm. CygX-N63 seems to actually correspond to a single object down to the 1\asec (1700 AU) spatial resolution. 
    }
     \label{isolated}
  \end{figure*}
   \begin{figure*}
   \centering
   \includegraphics[width=6.5cm,clip,angle=0]{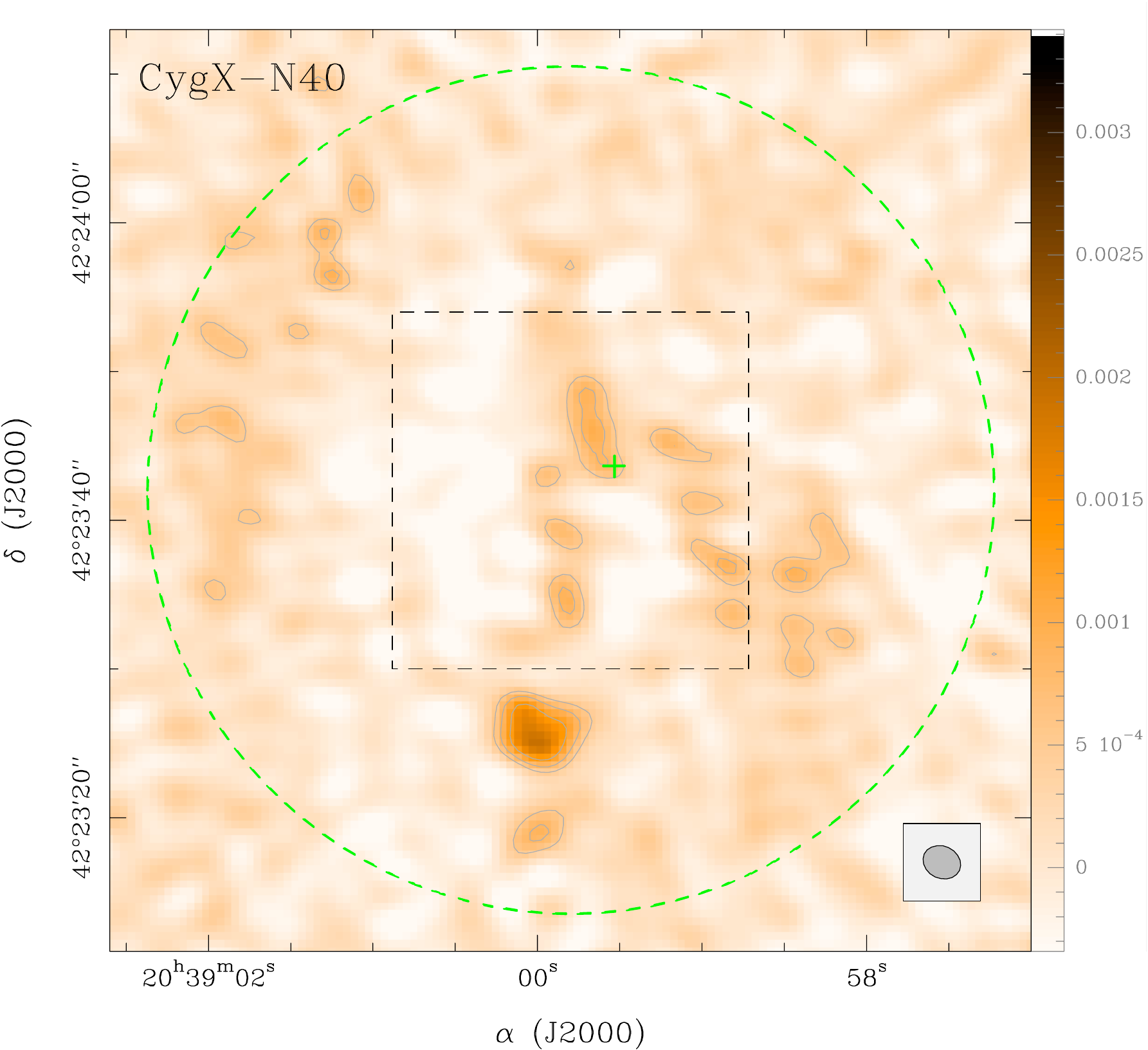}\hspace{0.23cm}
   \includegraphics[width=6.5cm,clip,angle=0]{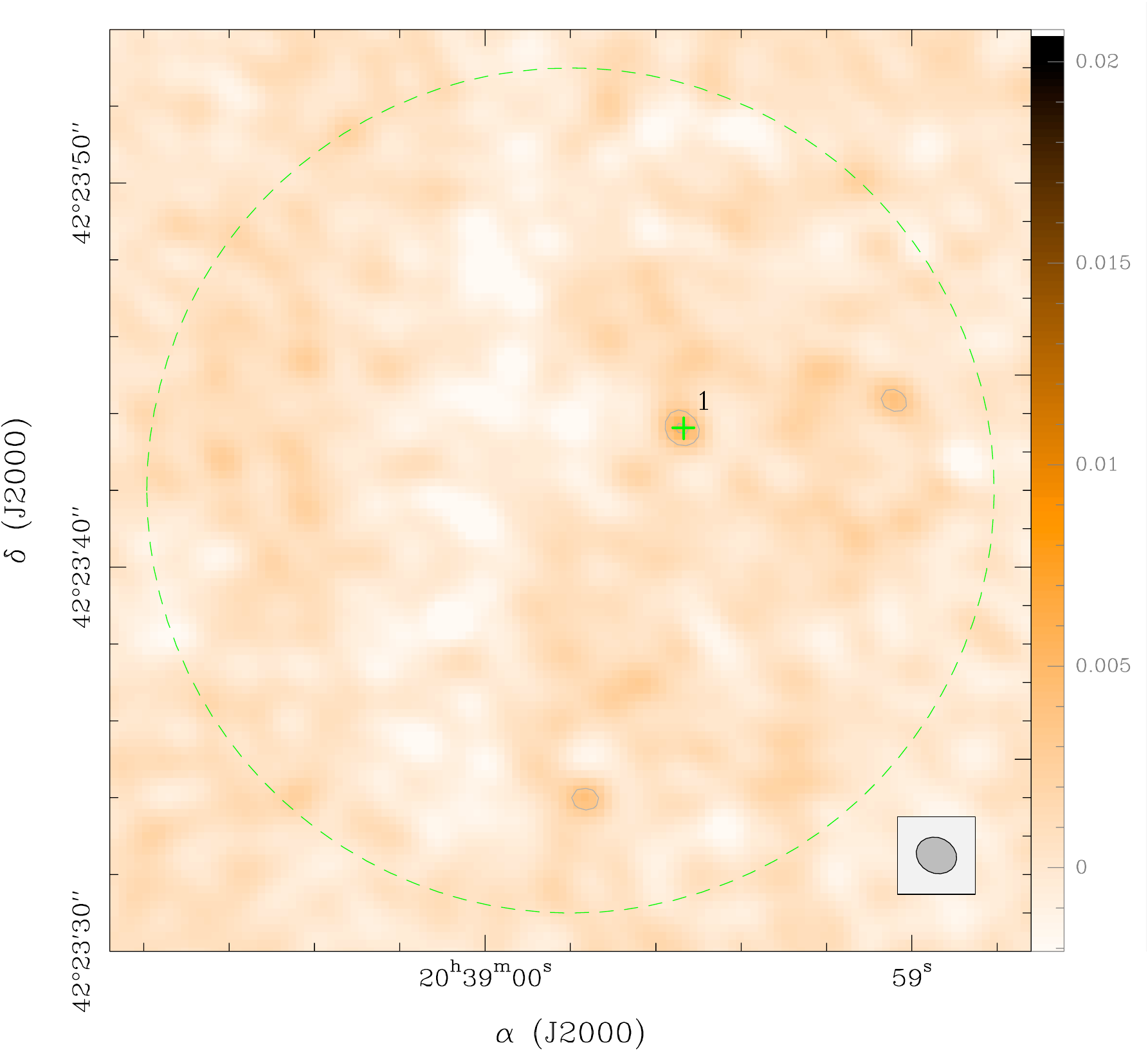}\\
   \includegraphics[width=6.5cm,clip,angle=0]{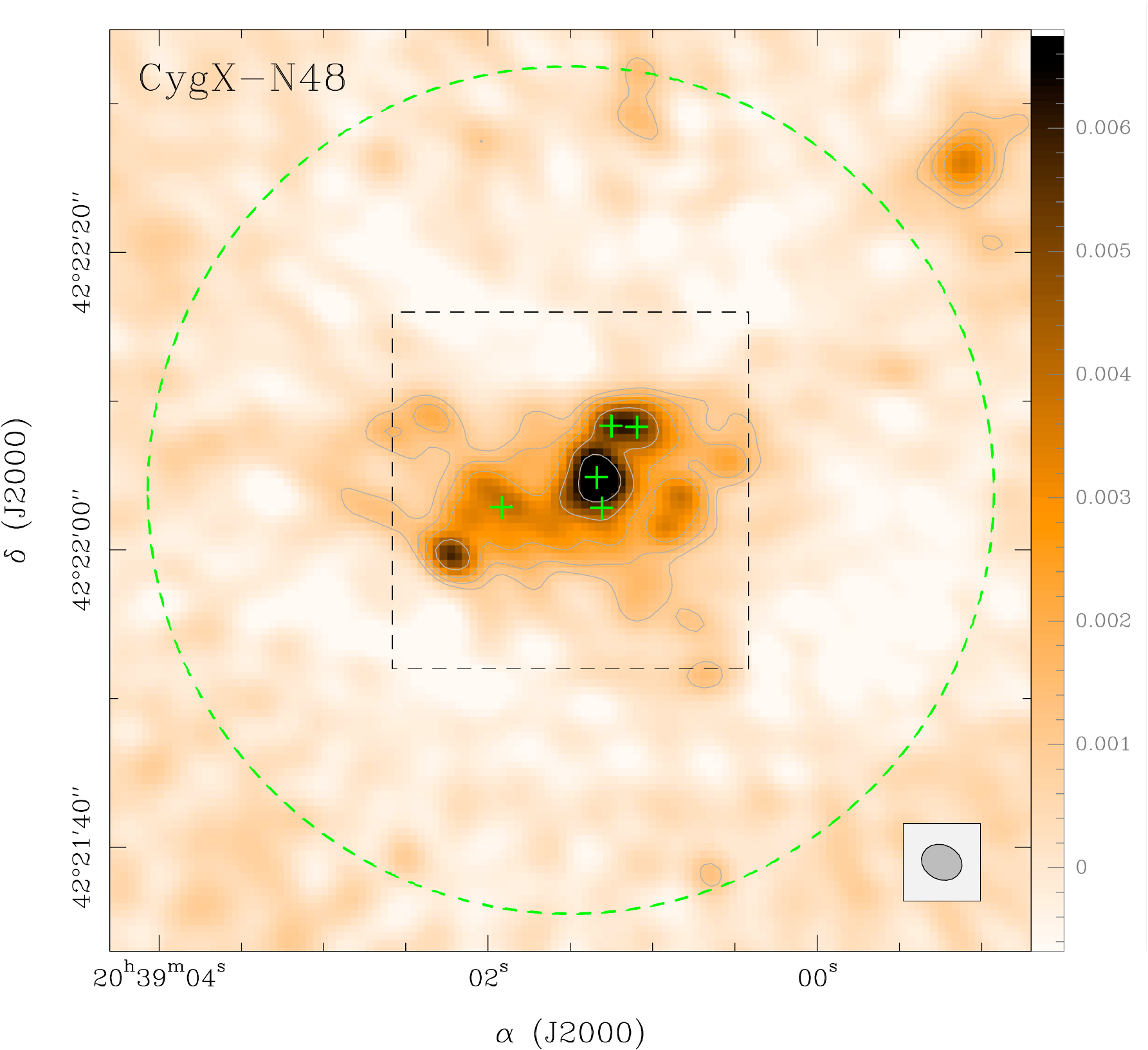}\hspace{0.23cm} 
   \includegraphics[width=6.5cm,clip,angle=0]{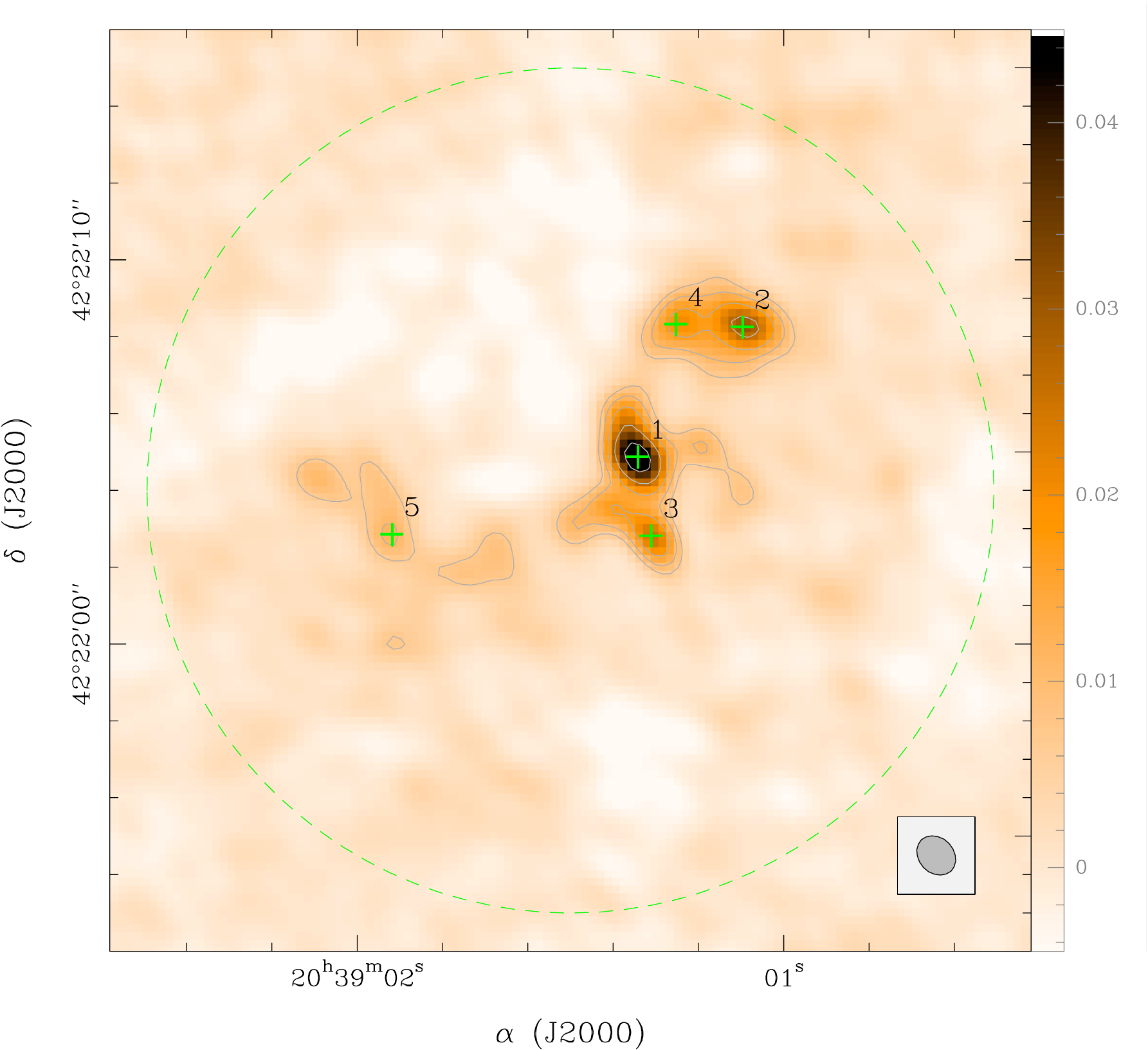}\\
   \includegraphics[width=6.5cm,clip,angle=0]{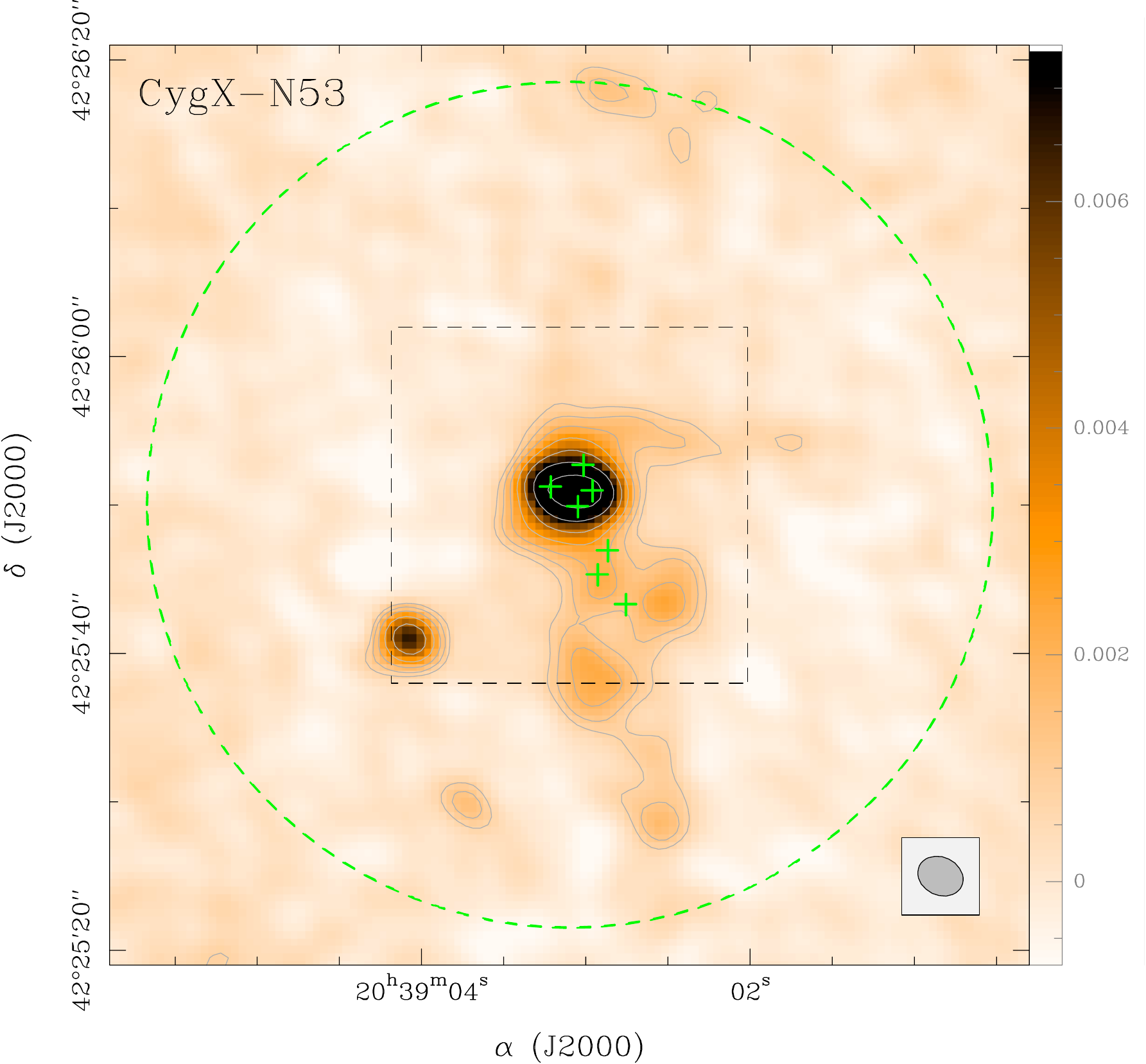}\hspace{0.23cm} 
   \includegraphics[width=6.5cm,clip,angle=0]{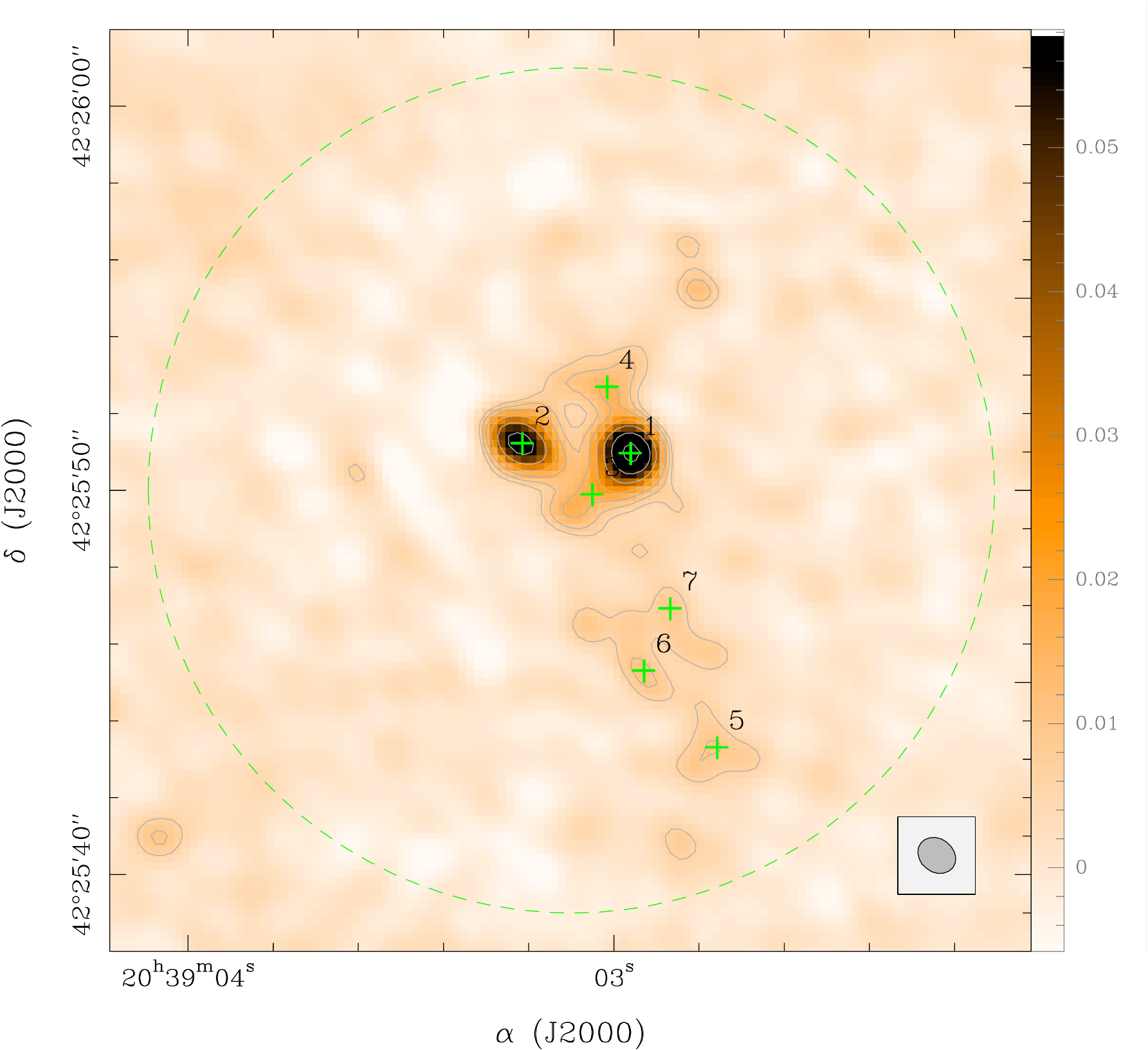}
   \caption{Same as Fig.~\ref{isolated} for the three MDCs located in the filament of DR21 (see text). Note that in contrast to the isolated MDCs (Fig.~\ref{isolated}), the cores show here diverse structures. CygX-N40 appears devoid of compact emission at both 3.5 and 1.3~mm. CygX-N48 has a complex and extended structure at 3.5~mm with two main cores in the central regions. The 1.3~mm emission resolve a few, probable compact, fragments in the two main cores. The bright peak at 3.5~mm in the South-East corresponds to the weak 1.3~mm source outside in the primary beam in the right panel. It is located at the edge of the nearby cluster associated with MSX6G81.7133+0.5589. CygX-N53 has a bright central core at 3.5~mm which splits into 4 smaller fragments at 1.3~mm in a way similar to the isolated MDCs. The 3.5~mm source in the SE is hardly detected at 1.3~mm due to the falloff in sensitivity of the primary beam (dashed circle).}
    \label{filament}
   \end{figure*}


\section{Results}
\label{results}

\subsection{The 3mm continuum maps}

The left panels in Figs.~\ref{isolated} and \ref{filament} display the dust continuum emission towards the six cores at 3.5~mm.
Strong 3$\,$mm emission is detected for all fields except CygX-N40. Despite a 
total mass of 106~\msol ~in Motte et al., CygX-N40 seems to be dominated by extended 
emission which is fully filtered out by the interferometer. We note that it has the lowest average density in our sample (see Tab.~\ref{global_properties}), and it is located in the large scale DR21 filament where it is possible that the observed emission is more elongated in the direction of the line-of-sight. This could have led us to mis-interpretate  CygX-N40 as a single MDC and to over-estimate the local density. It will therefore not be further considered for discussion. 
For the other fields, the emission contains bright compact emission peaks. They are structured into a few individual bright compact sources - 
except CygX-N63, which seems to be dominated by a single object. 
We note also that CygX-N48 appears less centrally concentrated than the other cores, and could be dominated by a large population of unresolved compact fragments forming a full cluster.
   \begin{figure*}
   \centering
 \includegraphics[angle=-90,width=13.0cm]{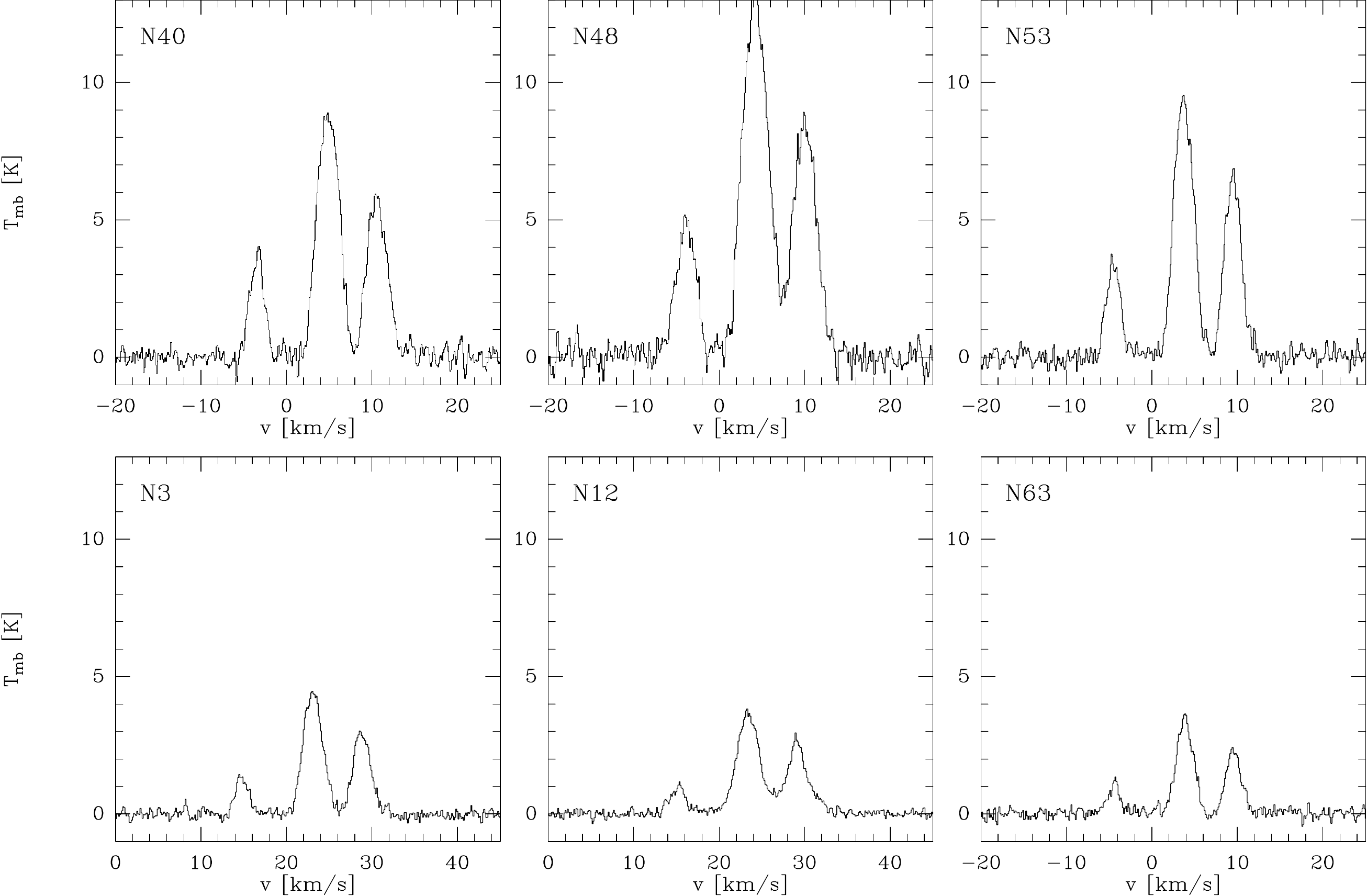}
   \caption{N$_2$H$^+$($1-0$) spectra towards the center of each MDCs observed with the IRAM 30$\,$m. Note that the velocity scale is given for the weakest line (left component) which is the only isolated component of the hyperfine structure. The other 5 components of the hyperfine structure are blended and form the 2 apparently brighter groups of lines.
  }
              \label{n2hp}%
   \end{figure*}

\subsection{The 1mm continuum maps}

The right panels in Figs.~\ref{isolated} and \ref{filament} display the dust continuum maps at 1~mm. For three of the cores (CygX-N3, CygX-N12, CygX-N53) the difference between low (3.5 mm) and high (1.3 mm) resolution maps is striking. While at a scale of 2\asec to 3\asec resolution, the emission mainly consists of extended, elongated structures, at the  $\sim\,$1\asec resolution of the 1.3~mm map the emission clearly splits into a number of individual usually spherical fragments. In contrast, CygX-N48 has a similar aspect than at 3.5$\,$mm with emission spread over the whole region and with compact sources which are not very spherical such as if it consists of partly non resolved individual objects. CygX-N63 seems to differ from the others since the emission is dominated at all observed scales by a single central object.  
A slight trend of having some distributions of aligned compact sources inside filamentary structures seen at 3.5$\,$mm can be noted. CygX-N3 and
CygX-N12 are composed of three to four aligned compact sources. CygX-N48 has a "S" shape emission which could be a 
filamentary structure too.
CygX-N53 displays an almost north-south ridge of emission which is aligned with the large scale orientation of the DR21 filament in which it is embedded. The direction of the axis between the two 
brightest sources is however oriented east-west, almost perpendicular to the main direction of the large scale emission.

\subsection{Source extraction at 1$\,$mm}
\label{extraction}

Our ultimate goal is to study the population of individual protostellar objects inside the MDCs of Cygnus X. We therefore extract fragments in the highest spatial resolution data maps  (1~mm maps).
To be systematic, we used the Gaussclump algorithm \citep{stutzki&guesten1990} adapted to continuum maps as described in  \citet{motte2007}
to extract the compact sources. Above a safe 7 sigma level, 24 sources have been
detected inside the six fields and 23 inside 5 fields (excluding CygX-N40), i.e. an average of almost 5 sources per
field. All sources are clearly identified already by eye-inspection in the maps.

The obtained list of fragments, ordered by increasing peak flux, is given in Tab.~\ref{list_sources}. The table gives the coordinates, peak fluxes, 2D gaussian sizes (FWHM), and integrated fluxes directly obtained from the Gausclump analysis. The integrated fluxes are converted into masses M$_{\rm PdBI}$, using a dust emissivity at 1.3$\,$mm of 1.0 cm$^2$g$^{-1}$  \citep{ossenkopf&henning1994}, a gas-to-dust mass ratio of 100, and a dust temperature of 20~K (formula (1) in \citealp{motte2007}). The de-convolved sizes (col.~8 in Tab.~\ref{list_sources}) correspond to the quadratic average of the de-convolved 2D sizes (FWHM) and are converted in AU for d=1.7 kpc. Average densities are calculated inside the projected region of radius  $R_{90}$ containing 90~\% of the mass (M$_{\rm PdBI}$) for a 2D gaussian distribution: 

 \begin{equation}
\label{r90}
    R_{90} = 1.95 \times {\rm FWHM}\,/\sqrt{ 8\,{\rm ln} 2}.
   \end{equation}

The masses M$_{\rm PdBI}$ obtained from the gaussian decomposition of the PdBI maps are good estimates of the masses inside the inner regions of the compact fragments. Their actual sizes and total masses can however not be easily derived from the present interferometric data which tend to filter out the extended envelopes (see Sect.~\ref{nature}). An estimate of the total masses  M$_{\rm tot}$ is given in Tab.~\ref{list_sources}. This is an extrapolation of M$_{\rm PdBI}$ inside a total
size of 3000 AU (FWHM). According to the detailed discussion given in Sect.~\ref{nature}, these masses should be seen as conservative estimates of the total masses
in contrast to M$_{\rm PdBI}$ which are only lower limits.

We take a temperature of 20 K for all mass derivations following \citet{motte2007}. This typical temperature has been recently confirmed by NH$_3$ measurements by \citet{wienen2008} using the Effelsberg telescope. The emissivity adopted \citep{ossenkopf&henning1994} is used in most studies of low-mass star-formation but also for high-mass protostellar objects (\citealp{andre2000}, \citealp{hunter2006}, \citealp{rathborne2007}, \citealp{schuller2009}) and should be adequate for dense and cold gas. Alternatively, some authors use the older values by \citet{hildebrand1983} usually in the case of warm objects for which the emissivities might be slightly different. These emissivities are 3 times smaller than the ones of \citet{ossenkopf&henning1994} in the millimeter range which is a significant difference, leading to masses $\sim3$ times larger than the ones we derived here. Any comparative analysis should therefore take into account these different assumptions on emissivities.

\begin{table*}
\begin{minipage}[t]{\columnwidth}
\caption{Global properties of the MDCs derived from single dish continuum and N$_2$H$^+$ line observations.}
\label{global_properties}
\centering
\renewcommand{\footnoterule}{}  
\begin{tabular}{cccccccccccc}
\hline \hline
Name  &   Mass          & Size & {\rm $< n_{\rm H_2}>$} & $\sigma_{\rm turb}$ & {\rm $M_{\rm vir}$} & {\rm $\lambda_{\rm Jeans}$} & {\rm $M_{\rm Jeans}$} & $\Sigma_{\rm MDC}$ & $\tau_{\rm ff}$   &  {\rm $\lambda_{\rm sep}$}  \\
           & [M$_\odot$]  & [AU] & [{\rm cm}$^{-3}$]          & [km.s$^{-1}$]               &  [M$_\odot$]                   &  [AU]                                           &  [M$_\odot$]                  &  [g.cm$^{-2}$]         & [yr]  &  [AU]     \\
\hline
 CygX-N3     &    84.0  &   20300  & 5.29~$\times$~10$^5$   & 0.98  &  91.0 & 8490 &   1.1  &   1.45  &   6.0~$\times$~10$^4$     &   6150  \\
 CygX-N12   &    86.0  &   20400 & 5.33~$\times$~10$^5$    & 0.89  &  75.5 &   8460 &   1.1  &   1.47  &   5.9~$\times$~10$^4$    &   3420  \\
 CygX-N40\footnote{CygX-N40 does not show any bright compact emission in the PdBI maps suggesting it has been misinterpreted as a MDC probably due to confusion on the line-of-sight towards the DR21 filament (see text for details). It is therefore not further consider for discussion.}   &  106.0  &   33200   & 1.52~$\times$~10$^5$  & 0.82 &  - &   - &  -  &   -  & -       &    -   \\
 CygX-N48   &  197.0  &   27300  & 5.10~$\times$~10$^5$   & 1.28  &  209 &  8640  &   1.1  &  1.88   &  6.1~$\times$~10$^4$      &   3830   \\
 CygX-N53   &    85.0  &   24000  & 3.24~$\times$~10$^5$   &  0.76 &  64.8 &  10850  &   1.4 &   1.05  &   7.6~$\times$~10$^4$   &   4180    \\
 CygX-N63   &    58.0  &   12300\footnote{CygX-N63  was actually not resolved out by MAMBO in \citet{motte2007}, and this size might actually correspond to an upper limit of the source size.}
   & 1.64~$\times$~10$^6$   &  0.72 &  29.8 & 4820  &   0.62  &   2.73  & 3.4~$\times$~10$^4$   &   -     \\
 \hline
\end{tabular}
\end{minipage}
\end{table*}

\subsection{N$_2$H$^+$ 1$\to$0 emission towards the MDCs}
\label{n2h+}

The obtained single spectra towards each MDC are shown in Fig.~\ref{n2hp}. We use these spectra to estimate the average turbulent velocity dispersion in the cores. N$_2$H$^+$ 1$\to$0 is well suited since it is usually optically thin and because its abundance drops drastically in the lower density, warmer outer layers which tend to dominate the line widths of most other molecular tracers. Since the profiles are not always perfectly gaussian, we did not use gaussian fits, but measured directly the widths at half maximum of the lines. For these measurements, we only use the isolated component of the hyperfine structure (HFS) which appears as the weakest line in Fig.~\ref{n2hp}. The other 5 components are blended and form the 2 brighter groups of lines. The resulting values converted into the gaussian 1D velocity dispersion $\sigma_{\rm turb}$ are given in Tab.~\ref{global_properties}.

The fits of the 6 components of the N$_2$H$^+$ 1$\to$0 multiplet can provide some indications of the opacity and of the "average" excitation temperature. The 6 relative distances and intensities of the HFS components of the 1$\to$0 line are given as input to a simultaneous gaussian fit to all components, assuming equal excitation temperature
, using the ``Method HFS'' feature of GILDAS\footnote{Grenoble Imaging and Line Data Analysis software}. 
These fits confirm that the lines are actually optically thin with values of $\tau$ between 0.03 and 0.2. Also the obtained excitation temperatures are typically around 20 K, conform with NH$_3$ observations.

\section{Discussion}
\label{discussion}

\subsection{Global properties of the MDCs}
\label{global}

The targeted MDCs are very dense cores with densities 10 times larger than in low-mass star-forming regions (see Tab.~4 in \citealp{motte2007}). They have total masses large enough to be single high-mass protostars or to form a whole cluster of stars. We review here their global properties in order to discuss which population of stars these cores are expected to form, based on the theoretical predictions of the usually promoted scenarios for high-mass star formation.

In the first columns of Tab.~\ref{global_properties}, we give the properties of the MDCs as directly derived from the MAMBO survey of \citet{motte2007}: the names, total masses, 1D deconvolved sizes, and average densities, as well as the 1D turbulent dispersion derived from the N$_2$H$^+$ line widths. The average densities are however not exactly the same as in 
\citet{motte2007}, since we derive them in slighly different volumes.  We average here the densities inside R$_{90}$  (Equ.~\ref{r90}), the radius which corresponds to 90$\,$\% of the mass for the 2D gaussian distribution used to extract the cores. 

In the second part of Tab.~\ref{global_properties}, we give the quantities which are believed to ultimately dictate the properties of the proto-stellar fragmentation and of the subsequent star formation: the virial masses $M_{\rm vir}$, the local Jeans lengths $ \lambda_{\mathrm{Jeans}}$ and Jeans masses $M_{\mathrm{Jeans}}$, the mean surface densities $\Sigma_{\rm MDC}$, and the free-fall times $\tau_{\rm ff}$. In addition the typical, observed separations $\lambda_{\rm sep}$ between the PdBI fragments is provided in the last column (see Sect.~\ref{mass} for details).
The  Jeans masses, Jeans lengths, and virial masses are calculated for the average conditions in the MDCs using the classical formulation: 

 \begin{equation}
      \lambda_{\mathrm{Jeans}} =
         \sqrt{ \frac{ \pi}{{\rm G} \rho_0} \frac{{\rm k} T}{m_{\rm{H_2}}} }
   \end{equation}

 \begin{equation}
      M_{\mathrm{Jeans}} =
         \frac{ 4 \pi}{3}  \rho_0 \left( \frac{ \lambda_{\mathrm{J}}}{2} \right)^3
   \end{equation}

 \begin{equation}
      M_{\mathrm{vir}} =
         \frac{ 5 \sigma_{\rm turb}^2  R}{G}  
   \end{equation}
   
 which numerically gives  
 
 \begin{equation}
      \lambda_{\mathrm{Jeans}} = 1.38 \times 10^4\,  \mathrm{AU} \, \left( \frac{T}{10\,\mathrm{K}}  \right)^{0.5}   \left( \frac{n_{\mathrm{H_2}}} {10^5 \, \mathrm{cm}^{-3}} \right)^{-0.5}   
   \end{equation}

 \begin{equation}
      M_{\mathrm{Jeans}} = 0.90\, \mathrm{M}_\odot  \, \left( \frac{T}{10\,\mathrm{K}}  \right)^{1.5}   \left( \frac{n_{\mathrm{H_2}}} {10^5 \, \mathrm{cm}^{-3}} \right)^{-0.5}  
   \end{equation}

 \begin{equation}
      M_{\mathrm{vir}} = 115.6\, \mathrm{M}_\odot  \,   \left( \frac{ R} {0.1 \, {\rm pc}} \right)    \left( \frac{ \sigma_{\rm turb}}{1\,{\rm km.s}^{-1}}  \right)^{2} 
   \end{equation}
The listed $\lambda_{\rm Jeans}$ and $M_{\rm Jeans}$ have been calculated for a temperature of 20 K (see Sect.~\ref{extraction}). In an earlier phase for the real initial conditions, the MDCs could have been colder, perhaps as down as 10 K such as what is observed in low-mass dense cores, leading to smaller Jeans lengths and masses. The virial masses are calculated for $R=R_{90}$ and  $\sigma_{\rm turb}$ is the 1D (average on the line-of-sight) velocity dispersion.
The values for  $\Sigma_{\rm MDC} = M / (\pi R^2) $ are calculated for the half-mass radius for proper comparisons with values promoted in \citet{mckee&tan2003}. 


The average densities in the MDCs are so high that the free-fall times are of the order of or smaller than 10$^5\,$yr. This is actually smaller than the statistical timescales for low-mass protostars derived in nearby star-forming regions.  The duration of the protostellar evolution, which corresponds to the sum of the lifetimes of Class 0 and Class I YSOs, is of the order of $2\times10^5\,$yr (\citealp[e.g.][]{greene1994}; \citealp{kenyon&hartmann1995}). The free-fall times are also roughly equal to the crossing times of the cores, estimated by dividing the sizes by the 3D velocity dispersions ($\sqrt{3}\times\sigma_{\rm turb}$), which are typically equal to $6\times10^4\,$yr. The MDCs are clearly self-gravitating since the ratios $M/M_{\rm vir}$ are all close to or even larger than 1 and up to 2 for CygX-N63. The turbulent support can therefore not prevent the cores from collapse.  The unknown magnetic field could delay the collapse but cannot avoid it. If they actually collapse in a free-fall time, the global infall rates will be of the order of a few $10^{-3}\,$\msol$/$yr which is high enough to overcome the radiation pressure barrier. These MDCs could therefore be in monolithic collapses leading to single massive stars. Their global properties (total masses, sizes, turbulent velocity dispersion, and mean surface densities) are conform with the initial conditions for the turbulence-regulated monolithic collapse proposed by \citet{mckee&tan2002}.
Alternatively, according to \citet{dobbs2005} for instance, such massive, turbulent cores should actually fragment  to form a whole cluster of stars. 
 
\begin{table*}
\begin{minipage}[t]{\columnwidth}
\caption{List of fragments and physical properties at 1.3~mm.}
\label{list_sources}
\centering
\renewcommand{\footnoterule}{}  
\begin{tabular}{lccccccccc}
\hline \hline
$\hspace{0.4cm}$Fragment &   \multicolumn{2}{c}{Coordinates} & Peak  &   FWHM\footnote{Not deconvolved FWHM of the 2D gaussian.}     &  Int.  &  M$_{\rm PdBI}$ &  Deconv.\footnote{Deconvolved FWHM in 1D (quadratic average of the 2 axis sizes) expressed in AU for d=1.7 kpc.} &   {\rm $< n_{\rm H_2}>$}\footnote{The average density is calculated inside the projected region containing 90 \% of the mass in a 2D gaussian assuming a spherical distribution.}
 &  M$_{\rm env}$ \footnote{M$_{\rm PdBI}$ extrapolated from the deconvolved size to a FWHM size of 3000 AU assuming a density profile in r$^{-2}$.} \\
$\hspace{0.5cm}$Name & R.A.      &   Decl.        & Flux   &    & Flux  &   & Size   &  &    \\
         &  (J2000)      &   (J2000)       & [{\rm mJy/beam}] & [$\,^{\prime\prime}\, \times \, ^{\prime\prime}\,$] & [{\rm mJy}] & [{\rm M}$_\odot$] &  [{\rm AU}]  & [{\rm cm}$^{-3}$] & [{\rm M}$_\odot$]\\
\hline
CygX--N3 MM1  & 20 35 34.63 & 42 20 08.8 &  37.3 & 1.27~$\times$~1.05 &  49.6 &   2.63 &  979 & 1.48~$\times$~10$^8$  &   8.06 \\
CygX--N3 MM2  & 20 35 34.41 & 42 20 07.0 &  30.2 & 1.34~$\times$~1.04 &  41.6 &   2.21 & 1050 & 1.01~$\times$~10$^8$  &   6.31 \\
CygX--N3 MM3  & 20 35 34.23 & 42 20 04.7 &  14.4 & 1.01~$\times$~1.00 &  14.5 &  0.77 &  $<$ 851 & $>$6.56~$\times$~10$^7$  &   2.71 \\
CygX--N3 MM4  & 20 35 34.55 & 42 20 00.3 &   8.7 & 2.01~$\times$~1.01 &  17.7 &   0.94 & 1720 & 9.65~$\times$~10$^6$  &   1.63 \\
\hline
CygX-N12 MM1  & 20 36 57.65 & 42 11 30.2 &  64.1 & 1.22~$\times$~1.12 &  92.6 &   4.91 & 1100 & 1.93~$\times$~10$^8$  &  13.36 \\
CygX-N12 MM2  & 20 36 57.51 & 42 11 31.2 &  54.7 & 1.46~$\times$~1.13 &  95.2 &   5.05 & 1420 & 9.23~$\times$~10$^7$  &  10.65 \\
CygX-N12 MM3  & 20 36 57.81 & 42 11 29.7 &  12.4 & 1.77~$\times$~1.19 &  27.4 &   1.45 & 1820 & 1.26~$\times$~10$^7$  &   2.39 \\
CygX-N12 MM4  & 20 36 57.45 & 42 11 32.4 &  11.7 & 3.11~$\times$~0.98 &  37.6 &   1.99 & 2460 & 7.00~$\times$~10$^6$  &   2.42 \\
\hline
CygX-N40 MM1  & 20 38 59.54 & 42 23 43.6 &   5.7 & 1.04~$\times$~1.00 &   5.9 &  0.31 &  $<$ 847 & $>$2.72~$\times$~10$^7$  &   1.11 \\
\hline
CygX-N48 MM1  & 20 39 01.34 & 42 22 04.9 &  53.2 & 1.58~$\times$~1.07 &  88.5 &   4.69 & 1400 & 9.05~$\times$~10$^7$  &  10.07 \\
CygX-N48 MM2  & 20 39 01.10 & 42 22 08.3 &  37.5 & 1.84~$\times$~1.28 &  86.6 &   4.59 & 1960 & 3.19~$\times$~10$^7$  &   7.01 \\
CygX-N48 MM3  & 20 39 01.31 & 42 22 02.8 &  25.6 & 1.65~$\times$~1.01 &  41.7 &   2.21 & 1360 & 4.63~$\times$~10$^7$  &   4.88 \\
CygX-N48 MM4  & 20 39 01.25 & 42 22 08.3 &  20.0 & 1.64~$\times$~1.14 &  36.8 &   1.95 & 1570 & 2.65~$\times$~10$^7$  &   3.72 \\
CygX-N48 MM5  & 20 39 01.92 & 42 22 02.9 &  12.8 & 2.34~$\times$~1.24 &  36.3 &   1.93 & 2330 & 8.04~$\times$~10$^6$  &   2.48 \\
\hline
CygX-N53 MM1  & 20 39 02.96 & 42 25 51.0 & 106.3 & 1.19~$\times$~1.09 & 151.0 &   8.00 & 1050 & 3.61~$\times$~10$^8$  &  22.80 \\
CygX-N53 MM2  & 20 39 03.22 & 42 25 51.2 &  71.1 & 1.39~$\times$~1.03 & 112.1 &   5.94 & 1230 & 1.66~$\times$~10$^8$  &  14.44 \\
CygX-N53 MM3  & 20 39 03.05 & 42 25 49.9 &  17.3 & 1.46~$\times$~1.45 &  40.2 &   2.13 & 1870 & 1.72~$\times$~10$^7$  &   3.42 \\
CygX-N53 MM4  & 20 39 03.02 & 42 25 52.7 &  16.2 & 1.44~$\times$~1.44 &  37.0 &   1.96 & 1840 & 1.66~$\times$~10$^7$  &   3.20 \\
CygX-N53 MM5  & 20 39 02.76 & 42 25 43.3 &  13.2 & 1.72~$\times$~1.60 &  39.8 &   2.11 & 2300 & 9.08~$\times$~10$^6$  &   2.75 \\
CygX-N53 MM6  & 20 39 02.93 & 42 25 45.3 &  11.8 & 2.21~$\times$~1.02 &  29.4 &   1.56 & 1980 & 1.06~$\times$~10$^7$  &   2.36 \\
CygX-N53 MM7  & 20 39 02.87 & 42 25 46.9 &   9.5 & 2.27~$\times$~1.07 &  25.3 &   1.34 & 2100 & 7.60~$\times$~10$^6$  &   1.92 \\
\hline
CygX-N63 MM1  & 20 40 05.39 & 41 32 13.1 & 274.3 & 1.39~$\times$~1.05 & 419.2 &  22.22 & 1210 & 6.64~$\times$~10$^8$  &  55.20 \\
CygX-N63 MM2  & 20 40 05.51 & 41 32 12.7 &  44.1 & 2.09~$\times$~1.44 & 139.4 &   7.39 & 2440 & 2.68~$\times$~10$^7$  &   9.09 \\
CygX-N63 MM3  & 20 40 04.97 & 41 32 22.5 &  37.8 & 1.63~$\times$~1.49 &  96.1 &   5.09 & 2060 & 3.05~$\times$~10$^7$  &   7.41 \\
\hline
\end{tabular}
\end{minipage}
\end{table*}

\subsection{Properties of low-mass, nearby  protoclusters}
\label{lowmass}

The earliest phases of low-mass star formation as observed in nearby star-forming regions correspond to the so-called pre-stellar cores and to Class 0 and Class I protostars. They appear to form in clusters in regions such as $\rho$~Ophiuchi, Perseus, Serpens, or Orion. Below we refer mostly to results obtained in $\rho$~Ophiuchi and Orion because these are reference regions, and our group has studied them using the same, homogeneous way to derive masses, and sizes like in Cygnus X. References for other nearby protoclusters can be found in recent reviews such as \citet{andre2000}, \citet{ward-thompson2007}, \citet{difrancesco2007}.

One of the important findings obtained in these regions is that the fragmentation of clouds and dense cores is probably at the origin of the stellar masses, and that the masses of objects at the beginning of the collapse accurately reflect the final stellar mass in absolute values, or at least recognizing that a large fraction (efficiency) end up in the star. \citet{motte1998}, followed by a large number of other works \citep[e.g.][]{testi&sargent1998, johnstone2000, nutter&ward-thompson2007}, could trace that the mass distribution of pre-stellar cores and Class~0 YSOs have the same shape than the IMF. The stellar masses seem to be determined already at the pre-stellar stage as a result of the fragmentation of dense cores. Inside these low-mass dense cores, the fragmentation efficiency seems to vary with the average density and is typically of the order of 5 to 20\% but has been found to reach 40\% in $\rho$~Oph A \citep[see Fig. 7 in][]{motte1998}. The fragmentation efficiency is defined as the fraction of mass inside the pre- and proto-stellar envelopes. On the other hand, the global star formation efficiency (hereafter SFE), which includes also the masses of the already formed stars, is on average more of the order of  30\% \citep{bontemps2001} at the scale of the dense cores in the $\rho$~Ophiuchi proto-cluster. The typical sizes of individual protostellar objects are of the order of a few 1000 AU (3000 AU in 
$\rho$~Ophiuchi and 5000 AU in Orion B for instance) and the separations between objects was found to be typically of 5000 AU in nearby proto-clusters.

\subsection{A population of dense, self-gravitating fragments}
\label{nature}

The MDCs in Cygnus X are clearly massive and self-gravitating. They are therefore expected to contain collapsing objects which could be true massive protostars. Our new PdBI maps at high resolution down to physical scales of only 1700 AU should reveal these protostellar objects. 

At the moderate spatial resolution of $\sim 3$\asec (5100 AU) at 3.5~mm (left panels in Fig.~\ref{isolated} and \ref{filament}), 4 out of the 5 MDCs consist of strong, centrally concentrated, emission emanating from single bright, sometimes elongated (CygX-N3 and CygX-N12), cores in the inner regions of the MDCs. But,  at the highest spatial resolution of $\sim 1$\asec (1700 AU) at 1.3~mm (right panels in Fig.~\ref{isolated} and \ref{filament}), the picture drastically changes. The inner cores of CygX-N3, CygX-N12, and CygX-N53 actually split into a number of roughly spherical fragments. A total of 11 of such fragments (3 to 4 per MDC) are found to have roughly similar sizes and separations. Only CygX-N63  stays 
single\footnote{The secondary source CygX-N63 MM2 does not seem to be convincingly separated from MM1. The comparison of the 3.5 and 1.3~mm maps shows that the global shape of the emission is unchanged with clear extensions of MM1 towards the south-west, east and north-west directions at both resolutions and wavelengths. The extraction of MM2 as an individual fragment seems more due to this east extension than to the existence of a secondary object.} 
in the 1.3~mm map.  The case of CygX-N48 appears slightly different with emission at 3.5~mm which is significantly less centrally condensed than the other 4 MDCs. The central emission already splits at 3.5~mm into at least two main cores instead of one in the other MDCs. At 1.3~mm, 4 possibly individual fragments could be resolved and extracted in two main cores, but these fragments stay less spherical and with smaller separations than in the other MDCs. The difference may be due to a less centrally condensed MDC, and to a possible larger population of weaker fragments which are not fully resolved due to crowding. Finally, CygX-N63 is single and thus counts for one additional fragment.
Altogether, we therefore consider that a population of 16 individual fragments has been recognized in the central regions of 5 MDCs. The remaining 7 sources listed in Tab.\ref{list_sources} are weaker sources usually in the outskirts of the MDCs which can be considered as more uncertain detections (some are not confirmed at 3.5~mm). 


We see that 2 out of the 23 individual fragments have M$_{\rm PdBI}$ larger than 8~\msol  (CygX-N53 MM1 and CygX-N63 MM1), and 5 additional ones larger than 4~\msol. The values M$_{\rm PdBI}$ actually measure more the masses of the inner regions of the fragments  than their total masses (see discussion below). Such masses are among the largest masses ever measured in molecular clouds, inside physical sizes of only $\sim$1000 AU. The resulting average densities are extremely high, up to more than $10^8\,$cm$^{-3}$ (measured in $R_{90}$) for 5 of them, and the resulting free-fall times are small, with an average value of only $3900$~yr for the inner regions of these 7 fragments. The typical velocity dispersion in the inner regions of the MDCs is not known. It is however reasonable to consider that it is not larger than the average dispersions $\sigma_{\rm turb}$ measured at the scale of the MDCs. In fact, the velocity dispersion is expected to decrease from large to small scales. If it would follow a classical Larson's law of index 0.5, the velocity dispersion, which is on average equal to $\sim 0.9\,$km.s$^{-1}$ at the 0.1~pc scale, would be 4.5 smaller at the 1000~AU scale, i.e. of the order of 0.2$\,$km.s$^{-1}$ which is close to the thermal velocity dispersion (0.27$\,$km.s$^{-1}$ for 20~K). Even in the extreme case there would be no decrease at all of the velocity dispersion at small scales, with a velocity dispersion of 0.9$\,$km.s$^{-1}$ in the inner regions, the virial mass for $R_{90}=830\,$AU (which corresponds to a size of 1000 AU in Tab.\ref{list_sources}) is 3.8~\msol ~which is still smaller than M$_{\rm PdBI}$ for these 7 most massive fragments.  We can therefore safely conclude that these fragments, or at least their inner regions, are self-gravitating, and are most probably collapsing objects. Another indication of the probable collapsing nature of these fragments is the general trend of the fragments to be spherical in the continuum maps such as if gravitation is shaping them. 

The typical separations between the fragments can be evaluated from the projected distances between the fragments. The de-projected\footnote{We use a de-projection factor equal to 1/sin(57.3\adeg)=1.19 which corrects for the unknown random orientation of the separation axis.}  average separations $\lambda_{\rm sep}$ between the closest neighbors of the 16 main fragments are given as the last column of Tab.~\ref{global_properties}. We note that the separations are of the order of 5000~AU, typical for values derived in nearby, low-mass proto-clusters. 

We have thus detected a population of self-gravitating fragments whose sizes and separations are similar to low-mass pre-stellar cores and protostars in nearby proto-clusters.  These fragments are most probably individual protostellar objects of similar nature, but more massive, than in nearby regions. Since the selected MDCs stayed undetected in the infrared, most of these objects should still be cold and young and should thus be more pre-stellar cores or Class~0 YSOs than Class~I YSOs. Also the selected MDCs are not associated with any strong radio centimeter sources from the literature. Three of them, CygX-N3, CygX-N12, and CygX-N63, were confirmed to be devoid of  3.6$\,$cm continuum sources at low levels, down to 0.1 mJy thanks to recent dedicated VLA observations (Bontemps et al. in prep).  No \uchii  ~region is therefore yet expected inside these cores. Also, several collimated CO outflows have been found to be driven by most of the massive fragments in our accompanying CO(2-1) PdBI data. These data will be presented in a forthcoming paper entirely dedicated to the nature and evolutionary status of the protostars.

\subsection{Total masses of the protostellar fragments}
\label{mass}

The sizes and masses of the fragments derived from the interferometer maps are affected by the spatial filtering due to the lack of information for the large scales (typically larger than 1/3 of the primary beam) but also due to the limitation of the deconvolution algorithm for limited UV coverages and noisy phases at high frequencies (1.3~mm) which affects also the intermediate spatial scales (sidelobes which cannot be cleaned). The stronger spatial filtering at 1.3~mm compared to 3.5~mm actually partly explains why the individual small scale fragments show up so clearly at 1.3~mm, the different spatial resolution does not explain all the differences. It is particulary visible for CygX-N3 and CygX-N53 where the 3~mm maps show strong, central, slightly extended emissions which do not seem to be entirely recovered at 1.3~mm where only the compact fragments are well detected. 
We therefore expect that the masses obtained at 1.3~mm are only valid for the smallest scales. We should thus regard the masses M$_{\rm PdBI}$  and sizes of the fragments as lower limits. 

In order to estimate what could be the true total masses of the fragments, we have extrapolated M$_{\rm PdBI}$  up to realistic sizes for such proto-stellar objects, and assuming a classical density profile in r$^{-2}$. The most reliable sizes are provided by what has been observed in nearby protoclusters. In the $\rho$~Ophiuchi protocluster, \citet{motte1998} derived an average size for 58 starless millimeter condensations of the order of 3000 AU (deconvolved FWHM) for masses in the range 0.1 to 1~\msol. In Orion B, the more massive condensations (masses up to $\sim $3~\msol)  were found to be slightly larger with an average size closer to 5000 AU (see Fig.~4 in \citealp{motte2001}). By comparing $\rho$~Ophiuchi and Orion, it seems that the cores tend to be larger in size when they are more massive. 
In Cygnus X, and for higher mass pre-stellar and protostellar cores, we could therefore expect even larger sizes for the collapsing cores. This is also what is predicted in \citet{mckee&tan2003}. 
We therefore adopt here the conservative, smaller size of 3000 AU like in $\rho$~Ophiuchus to extrapolate the PdB masses, and to derive the values M$_{\rm tot}$ given in Tab.~\ref{list_sources}.


\subsection{Progenitors of OB stars ?}
\label{obstars}

The obtained masses M$_{\rm tot}$ then represent the gravitationally bound masses associated with each proto-stellar object. The bulk of these masses is expected to end up in the final star. Only the feedback from the outflow can decrease the global efficiency. Since outflows are well collimated during the earliest phases of the accretion phases (Class 0s) that corresponds to the phase for which most of the mass is accreted (e.g. \citealp{bontemps1996}) it is reasonable to assume that the efficiency stays rather high even considering outflows. The outflows from high-mass Class 0 YSOs are less known and could perhaps be less collimated \citep[e.g.][]{arce2007}. To be conservative, we thus assume here a rather low efficiency of 70~\% \citep[e.g.][]{bontemps2001}. Using this efficiency, we derive that all fragments with initial mass larger than 11.4~\msol will form a star more massive than  8~\msol. This is the case for 4 fragments. But, of course, statistically, we expect that the observed protostars have already accreted a fraction of their final mass. With a typical fraction of 50~\% of the mass already accreted, the statistical limit between intermediate mass protostars and high-mass protostars is then equal to 5.7~\msol. This is the case for 5 additional fragments. It is therefore a total of 9 fragments which could be precursors of ionizing OB stars (earlier than B3, $M_\star > 8\,$\msol). This estimate is based on statistical arguments and is therefore subject to uncertainties. The total number of O stars in the whole Cygnus X region is necessary larger than the 100 O stars recognized in the 2 Myr old Cygnus OB2 association (see \citealp{knodlseder2000}). Assuming 50~\% more O stars spread over the whole Cygnus X complex, we expect 150 O stars in total, and of the order of 750 ionizing OB stars (assuming a \citealp{kroupa2001} IMF) formed during the last 2 Myr. For a constant star formation rate, the 9 OB star precursors in the MDCs of Cygnus X would then correspond to a timescale of $2.4 \times 10^4\,$yr  quite typical for Class~0 YSOs (\citealp{andre2000}). This time is compatible with the free-fall times of Tab.~\ref{global_properties} and corresponds to 4 \msol ~(50~\% of the lowest stellar mass of OB stars) accumulated at a rate of $1.7\times10^{-4}\,$M$_\odot$/yr.





   \begin{figure}
    \includegraphics[width=6.0cm,clip,angle=270]{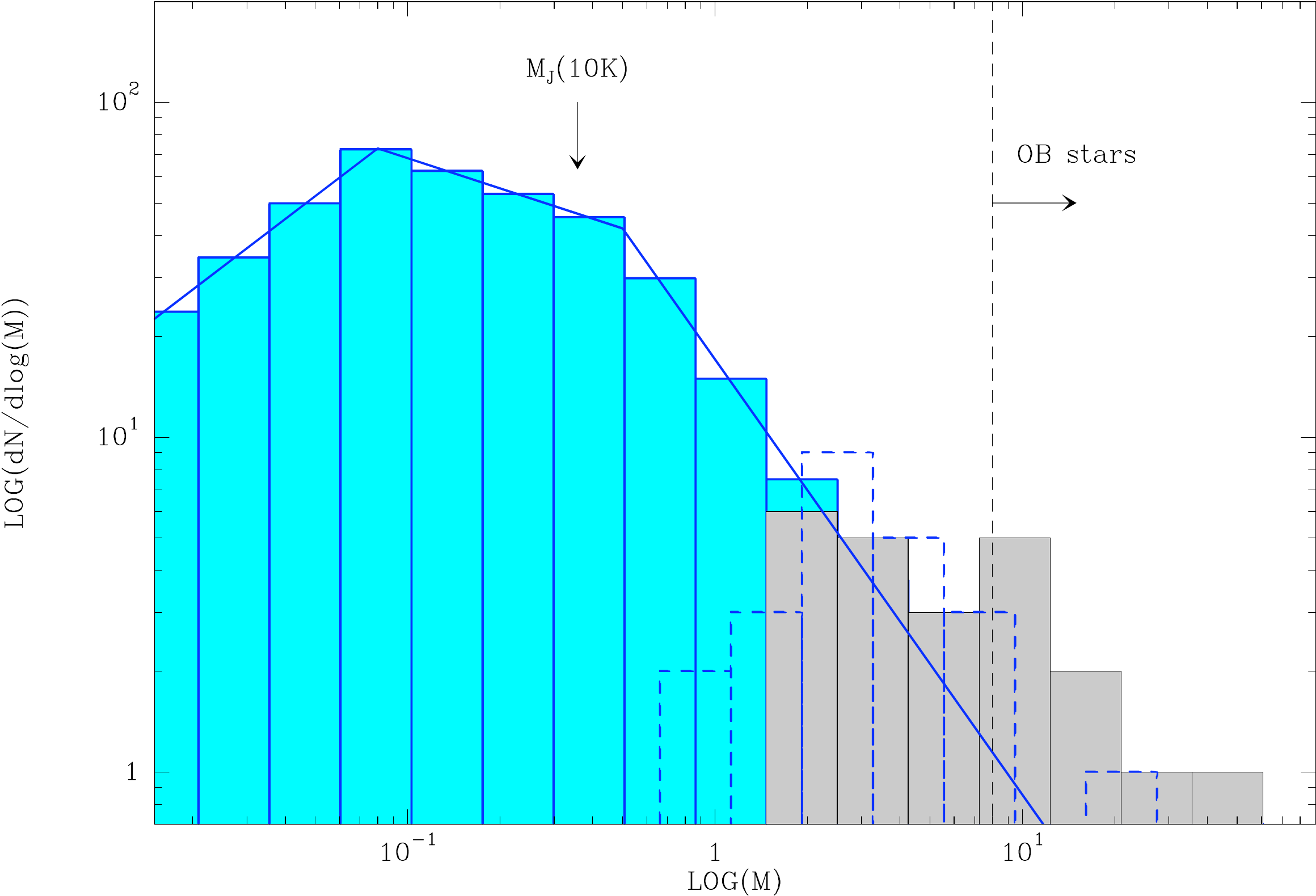}\hspace{0.23cm} 
    \caption{Mass function of the 23 fragments in the 5 MDCs considered in the discussion (grey histogram) using the values M$_{\rm tot}$. For comparison the global stellar mass function is shown for a global star formation efficiency of 30~\% of the total mass of 510$\,$\msol for the  5 MDCs, and using the \citet{kroupa2001} IMF for $M_\star$ between 0.01 and 120$\,$\msol (blue histogram in the background). The low-mass fragments in the MDCs cannot be detected but it seems that already the detected massive objects are over-massive compared to what a normal IMF would predict. Even in the extreme view of using the highly filtered masses M$_{\rm PdBI}$, there is still a slight excess of massive objects (dashed blue histogram).
    }
     \label{mf}
  \end{figure}

\begin{table*}
\begin{minipage}[t]{\columnwidth}
\caption{Properties of the star formation activity in the MDCs, and predictions assuming a normal IMF with SFE=30~\%.}
\label{obs_properties}
\centering
\renewcommand{\footnoterule}{}  
\begin{tabular}{ccccccccc}
\hline \hline
Name  &   Mass          & {\rm $M_{\rm tot}^{\rm PdBI}$} & {\rm $M_{\rm tot}^{\rm frag}$}  & {\rm $M_{\rm tot}^{\rm MYSO}$} &   {\rm $M_\star^{\rm 30\%}$}  &   {\rm $N_\star$}\footnote{It includes all stars and brown dwarfs from 0.01 to 120~\msol among which 63~\% are stars.}  &  {\rm $M_\star^{\rm max}$} &  {\rm $M_{\rm obs}^{\rm max}$} \\
            & [M$_\odot$]  &   [M$_\odot$(\%)] &   [M$_\odot$(\%)]   &   [M$_\odot$(\%)]  & [M$_\odot$]  &                               &   [M$_\odot$] &   [M$_\odot$] \\
\hline
 CygX-N3     &    84.0  & 7.2(8.6)  & 18.7(22.3)   &  14.4(17.1)  &    25.2   &      66  &  2.9  &  8.1  \\
 CygX-N12   &    86.0  & 17.3(20.1) & 28.8(33.5)  &  24.0(27.9) &    25.8   &      68  &    3.0   &  13.4  \\
 CygX-N48   &  197.0  & 26.1(13.3)& 28.2(14.3)   &  22.0(11.2) &    59.1   &      156  &  5.6   &  10.1  \\
 CygX-N53   &    85.0  & 28.5(33.6) & 50.9(59.9)   &  37.2(43.8)  &    25.5   &      67  &  2.9    &  22.8  \\
 CygX-N63   &    58.0  & 35.5(61.2)  & 64.3(111)   &  64.3(111)   &    17.4   &      46  &  2.2  &  55.2   \\
\hline
 Total          &    510.0  & 114.6(22.5) & 190.9(37.4)   &  161.9(31.7)   &    153.0   &    403  &  11.3   &  55.2  \\
 \hline
\end{tabular}
\end{minipage}
\end{table*}

\subsection{Comparison with previous studies of MDCs}
\label{previous}

A number of high-mass star-forming regions has already been investigated at high spatial resolution with (sub)millimeter interferometers, mostly with the IRAM PdBI and the SMA. But only a few studies were dedicated to the earliest phases of star formation and to regions of 0.1~pc or smaller. 
In fact, the presently available millimeter interferometers are limited in spatial resolution and only the most nearby MDCs can provide access to the smallest physical scales required to reach the individual proto-stellar objects. A few nearby HMPOs from the \citet{sridharan2002} sample were observed with the PdBI  \citep[see for instance][]{beuther2002c, leurini2007}, while a sample of not too distant IRDCs is presented in  \citet{rathborne2007}. These MDCs were all found to be sub-fragmented, with the presence of a few high-mass fragments or condensations which are good candidates to be direct precursors of OB stars. A mass function of intermediate mass fragments could even be derived by  \citet{beuther&schilke2004} for IRAS19410+2336, where it is recognized that the core mass function has roughly an IMF-like shape towards the high-mass regime too. However, the detected high-mass condensations were not so clearly recognized as compact, spherical fragments. Detections show only elongated emission distributions and weak peaks, moreover, many at the limit of the detection level (e.g. the PdBI  1~mm maps of \citealp{rathborne2007}). The better quality of our maps compares more with the results obtained by \citet{peretto2007} for a nearby MDC ($d=800\,$pc), NGC2264-C which has a size of 0.3 pc and contains two intermediate mass Class~0 YSOs at its center. The here presented statistical analysis of several MDCs in the same molecular complex enables to establish a core mass function with 9 objects in the high-mass regime, and discuss the statistical properties of the fragmentation of MDCs.

\subsection{Do massive fragments result from a turbulence regulated evolution?}
\label{protocluster}

The present observations clearly demonstrate that most of the MDCs are sub-fragmented and should therefore be classified as proto-clusters while they were found to have exactly the required masses, surface densities, and line widths to be single precursors of OB stars in the turbulence-regulated view of \citet{mckee&tan2002} (see Sect.~\ref{global}). 
This would be acceptable if thes MDCs would, at the end, contain only low-mass protostars. But this is not the case, these proto-clusters actually contain massive fragments, which  have the potentiality to form OB stars. These fragments have the same separations, and probably roughly the same sizes than in nearby, low-mass proto-clusters, making them significantly more compact and dense than the \citet{mckee&tan2002} cores.  Such smaller sizes seem difficult to reconcile with the slow turbulence regulated evolution to turbulence supported cores which have to be in equilibrium with their surroundings. In this scenario, the size of the cores results from the Larson scaling laws of turbulent clouds.

Only the most massive fragment, CygX-N63 MM1, seems to be a good candidate to be a turbulence regulated massive core. Interestingly enough, it is also the most massive single object of the sample. With a size of 0.06 pc (FWHM) and therefore a half-mass radius of $\sim0.03$pc, it has a size very close to the core radius predicted in \citet{mckee&tan2003}. On the other hand, this may just indicate that CygX-N63 is following the classical scaling laws and did not fragment further for some unknown reason, or that it will sub-fragment in the near future. Also some further investigations are required to establish in which kinematic state this core is. In fact, the line width in CygX-N63 is actually rather small for such a massive core in the quasi-static evolution view of  \citet{mckee&tan2002}. It leads to an exceptionally  large M/M$_{\rm vir}$ ratio of 2.7. The line width is only 2.7 times larger than the thermal width at 20 K. To understand the origin of such a core, it is therefore necessary to further investigate its properties, and especially its kinematics, and the nature of the observed velocity dispersion to establish, for instance, whether or not it is a true isotropic turbulent dispersion.

\subsection{IMF and the core mass function of the fragments}
\label{cmf}

In the simplified view that the MDCs are proto-clusters, they should fragment into a population of proto-stellar objects with a mass function which mimics the IMF of stars.  With a typical SFE of 30$\,$\%, a 100$\,$\msol$\,$ MDC would form of the order of 80 stars (including $\sim 30$ brown dwarfs) with a maximum stellar mass of $\sim 3.3\,$\msol$\,$ for a Kroupa IMF \citep{kroupa2001}. The total number of stars $N_\star$ and the mass of the most massive star $M_\star^{\rm max}$ expected to be formed in each MDC, and globally for the 5 MDCs, are given in Tab.~\ref{obs_properties}. For comparison, the observed maximum masses of the fragments in each MDC are also listed. It is clear that they are all systematically larger than what the IMF would predict. The mass function of the 23 detected fragments in the all 5 MDCs in Tab.~\ref{list_sources} is plotted in Fig.~\ref{mf} in comparison with the expected global IMF for the 5 MDCs. Our PdBI data cannot detect any low-mass fragments, and the completeness limit is high, probably close to 3 to 4$\,$\msol$\,$ for 4 of the 5 MDCs, and even higher for CygX-N63 for which it would be difficult to recognize a fragment of less than $\sim 10\,$\msol$\,$ (CygX-N63 MM2 with a mass of 9.1$\,$\msol$\,$ is actually a highly uncertain detection). This figure illustrates that there is a global excess of massive fragments compared to normal low-mass proto-clusters which mimic the IMF. Even in the very un-probable case that the fragments would actually not be larger than the sizes derived with Gaussclump on the interferometric maps, the mass function of $M_{\rm PdBI}$ is still in excess to the IMF (dashed histogram in Fig.~\ref{mf}). 

Globally for the 5 MDCs, with a SFE of 30$\,$\%, only one OB star should statistically be formed with a mass of 11.3$\,$\msol. The second most massive star should be a 6.8$\,$\msol ~star. It seems therefore clear that the most massive IR-quiet MDCs of Cygnus X are the location of the formation of a large fraction of high-mass proto-stellar objects which are likely candidates to form OB stars in excess to a normal IMF-like mass function. We further discuss in the following sections the important implications of this result.

\subsection{Do gravo-turbulent fragmentation plus competitive accretion can explain the observations?}
\label{fragmentation}

Due to the local, low values of the Jeans mass, the MDCs are actually expected to fragment and form clusters of protostars  \citep{dobbs2005} which may later collect more mass in the central regions by competitive accretion \citep{bonnell&bate2006}. In the case of strong radiative warming from already formed massive stars, the local Jeans mass could be larger and the level of fragmentation (number of fragments) could be reduced and could therefore form higher mass fragments \citep[e.g.][]{krumholz&bonnell2007}. The local temperatures are however not high enough (yet?) to have Jeans masses of the order of 10$\,$\msol$\,$ (see Tab.~\ref{global_properties}). 
Note also that the turbulent fragmentation has difficulties in forming high-mass fragments leading to core mass functions with a deficit of high-mass cores \citep[see discussions in][and references therein]{bonnell2007}. An additional process, such as competitive accretion, is required to explain high-mass stars, and to fully explain the observed IMF in this view.

The observed fragmentation of the MDCs could be seen as a support for the gravo-turbulent fragmentation scenario. The fraction of mass in the high-mass fragments, and the level of fragmentation (number of fragments) in the MDCs are however clearly not compatible with the pure gravo-turbulent fragmentation which would lead to a whole cluster of mostly low-mass objects (see previous section). The most striking examples are CygX-N12,  CygX-N53 and CygX-N63 which have 28, 44 and 100~\% of their total mass, respectively, in their two most massive fragments (see Tab.~\ref{obs_properties}) while only a few percents would be expected. In the context of the Bonnell and coll. view, a large fraction of the Cygnus X MDCs would already be in the stage after an intense competitive accretion to explain such high-mass objects. This might be surprising since the cores are still cold and therefore young. On the other hand, the high fragmentation efficiency (fraction of mass in dense fragments; see next section) indicates that the proto-clusters are already dominated by gravity of the proto-stellar objects as required for competitive accretion. But it would have time to work only if the crossing time in the inner region is shorter than the 
evolutionary time. If the evolution is dominated by the gravity, this evolutionary time is close to the global free-fall time which is typically smaller than the crossing time at the scale of the cores (see Sect.~\ref{global}). So to have a significant competitive accretion, the crossing time should be significantly smaller in the inner regions with typically an increase or at least a conservation of the velocity dispersion at the smallest scales. This has to be further investigated. Alternatively, the bias of the fragmentation towards high-mass fragments could  be primordial if, for instance, another support such as the magnetic field could prevent from a high degree of fragmentation as it was already obtained using MHD numerical codes for the collapse of low-mass pre-stellar cores \citep{hennebelle&teyssier2008}.

\subsection{High fragmentation efficiencies in MDCs}
\label{sfe}

To further discuss the resulting fragmentation in the MDCs, we define the fragmentation efficiency as the fraction of the total mass in the MDCs which is in self-gravitating fragments. Since the targeted MDCs are cold and young, we expect that most of the mass of the formed stars is still in the protostellar envelopes/fragments and that the fragmentation efficiency can then be seen as a proxy for the SFE in the MDCs. The degree or level of fragmentation discussed in the previous section is more related to the number of fragments, and could be referred to the multiplicity of the fragmentation. It is interesting to note that the fractions of the single dish emissions recovered by the interferometer are exceptionally high in our sample, and especially for CygX-N12, CygX-N53 and CygX-N63. This indicates that a large fraction of their mass is in small-scale structures. Since the small scale structures are expected to consist of self-gravitating fragments, we believe that this indicates a high fragmentation efficiency. 
To quantify this effect, we have measured the fraction of the single-dish fluxes recovered by the interferometer by integrating the emission above 3$\,\sigma$ associated for all sources detected at a 5$\,\sigma$ level with Gaussclump in the 1.3~mm maps. The obtained values, expressed in terms of masses ($M_{\rm tot}^{\rm PdBI}$) and related fractions are listed in Tab.~\ref{obs_properties}. They range between 10 and 20~\% for 3 of the 5 MDCs which are rather typical values for such massive cores but reach the exceptional values of 34 and 61~\% for CygX-N53 and CygX-N63. Alternatively,  the fragmentation efficiency is simply evaluated as $M_{\rm tot}^{\rm frag}$ by summing up the masses of the fragments listed in Tab.~\ref{list_sources} ($M_{\rm tot}$). This second estimate $M_{\rm tot}^{\rm frag}$ is more indirect since it is based on the extrapolated masses M$_{\rm tot}$ but it may approach closer the true fragmentation efficiencies as it is corrected for the spatial filtering of the interferometer. We see that for CygX-N3 and CygX-N46, the efficiency is below 25~\% and is therefore not exceptionally high. CygX-N12 is intermediate with an efficiency which migh be close to 30~\%. In contrast, CygX-N53 and CygX-N63 are clearly above 30~\%, indicating that the mass of these 2 MDCs is dominated by dense fragments, and that they may convert a large fraction of their mass into stars.
The case of CygX-N63 can be seen, in this context, as the extreme case where the fragmentation was so efficient that all the mass is in a single centrally concentrated object\footnote{The obtained $M_{\rm tot}^{\rm frag} = 64.3\,$\msol$\,$ is larger than the 58$\,$\msol$\,$ obtained from the gaussian fit of the MAMBO data in \citet{motte2007}, but only slightly, and is therefore compatible with all the mass of the MDC in a single object.}. But interestingly enough, the MDCs with the highest SFEs are also the ones for which most of the mass is in a few massive fragments with virtually 100$\,$\% of the mass in a single massive object for CygX-N63, 44$\,$\% of the total mass of CygX-N53 in only two high-mass fragments (27$\,$\% in MM1 and 17$\,$\% in MM2), and still 28$\,$\% for CygX-N12 in two fragments. We have listed in Tab.~\ref{obs_properties} as $M_{\rm tot}^{\rm MYSO}$  the total mass in the high-mass fragments (the 9 precursors of OB stars discussed in Sect.~\ref{obstars}) for each MDC. 


   \begin{figure}
    \includegraphics[width=8.0cm,clip,angle=0]{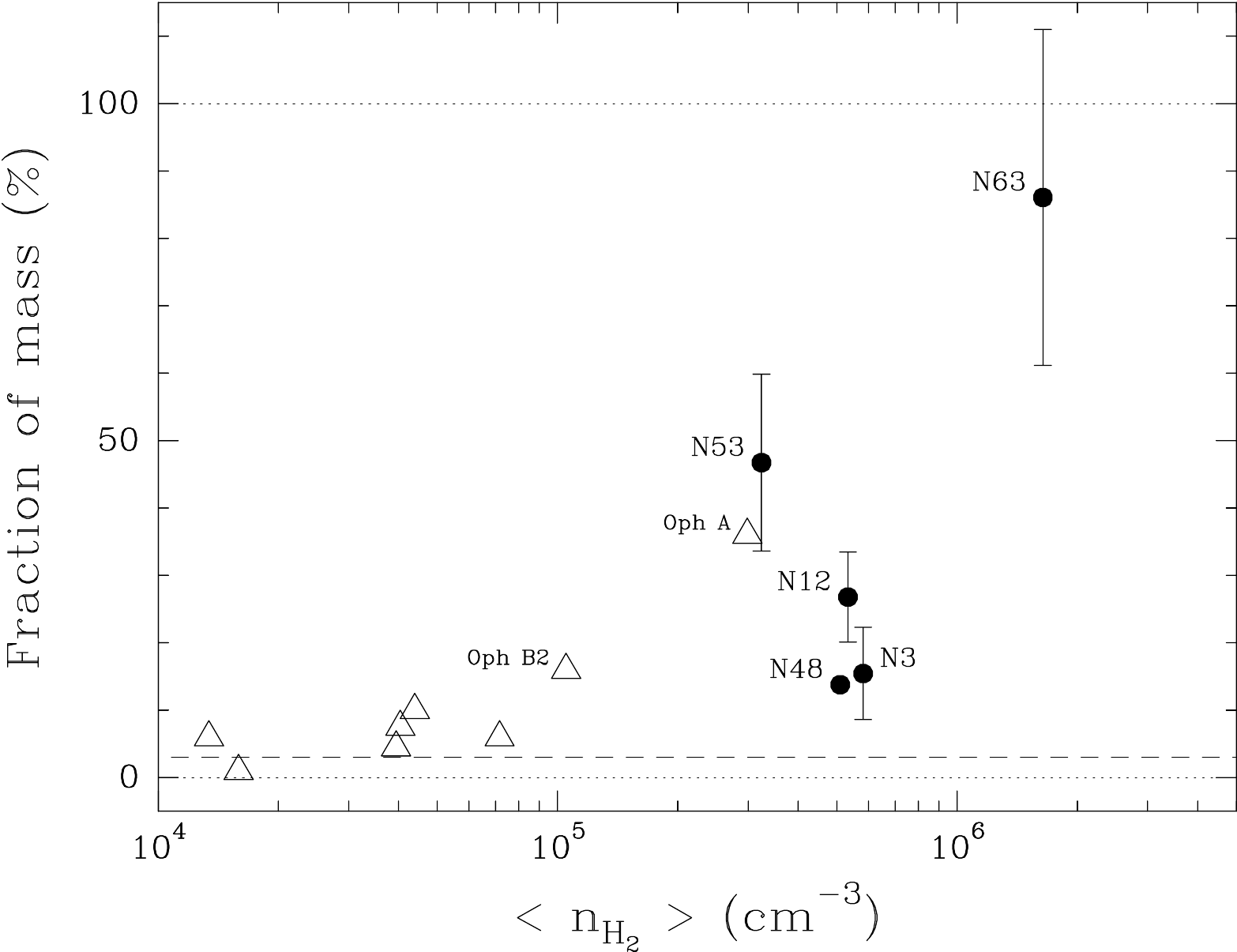}\hspace{0.23cm} 
    \caption{Fragmentation efficiency as a function of the average densities for dense cores of similar sizes (FWHM $\sim 0.1\,$pc) but of different masses in Cygnus MDCs (filled symbols) and in $\rho$~Ophiuchus \citep[open triangles; ][]{motte1998}. A transition from low efficiency cores to possible, single collapsing cores is observed from a few 10$^4$ to a few 10$^6\,$\cmc. The displayed ranges (uncertainty bars) for the Cygnus cores correspond to the range between $M_{\rm tot}^{\rm PdBI}$ and $M_{\rm tot}^{\rm frag}$ from Tab.~\ref{obs_properties}. To help the clarity of the plot, the displayed location of CygX-N3 corresponds to its average density shifted by a factor of 1.1.    }
     \label{sfe}
  \end{figure}

Also we note that the most extreme MDC, CygX-N63, has the highest average density (see Tab.~\ref{global_properties}). \citet{motte1998} obtained, in $\rho$~Ophiuchus, a similar trend of an increase of fragmentation efficiency as a function of density (see their Fig.~7).
The fragmentation efficiencies of the MDCs are displayed in Fig.~\ref{sfe} as a function of the average densities derived in $R_{90}$ from Tab.~\ref{global_properties}. For comparison, the values obtained for similar size, but less massive dense cores by \citet{motte1998} are plotted as open triangles. We have re-evaluated the densities to the way we derived them in the present paper for a proper comparison. It is striking that the Cygnus MDCs are clearly confirming the trend suggested by the efficiencies in Oph A and B2  that the highest density cores have a larger fraction of their masses inside pre-stellar and protostellar objects. 
The origin of these high efficiencies has to be further investigated. We can however note that the higher the average density, the shorter the free-fall time is with the possible consequence that a global collapse at the scale of the core is expected for the highest density cores. In this simplified view, the clear trend in Fig.~\ref{sfe} could correspond to the transition from low density cores in the standard, large scale equilibrium between turbulent, magnetic pressures and gravity and dense cores dominated by their self-gravity leading to a virtually 100~\% efficiency in single, collapsing protostars. 



\subsection{Primordial mass segregation in proto-clusters?}

No matter what is the origin of the most massive fragments in the MDCs of Cygnus X, it is clear that the population of protostars in the center of the observed MDCs is biased towards high-mass objects. This is at least certainly true for CygX-N12, CygX-N53 and CygX-N63. It is a representative result since it was obtained from a complete sample of dense cores for the whole Cygnus X region. This clear excess of high-mass protostars in the center of a few MDCs in Cygnus X 
is actually what is expected to explain the primordial mass segregation of stellar clusters. The young stellar clusters, such as the Orion Nebula Cluster, tend to have their highest mass members only in their central, densest regions (\citealp{hillenbrand&hartmann1998}). For the youngest and not too rich clusters, the observed mass segregation can not be due to early dynamical relaxation and is understood as the evidence for a primordial mass segregation originating from the star/cluster formation process itself (see also \citealp{bonnell2007} and references therein).
Our new results on the MDCs of Cygnus X therefore demonstrate the existence of such a primordial mass segregation, at least in some MDCs/clusters.

If the MDCs correspond to the mass segregated, inner regions of the proto-clusters, where and when do the low-mass stars form? 
Inside the most extreme MDCs of Cygnus X (CygX-N12, CygX-N53 and CygX-N63) which are clearly mass segregated, it seems that there is not enough mass left to form a large population of low-mass stars to compensate for the observed excess of high-mass stars. So some surrounding gas has to be collected in the near future from some lower density gas around the MDCs to form the low-mass counterparts. 
We note however that this star-forming gas cannot be located too far from the center of the MDCs in order that the formed stars can incorporate the final cluster. To be dynamically linked to the cluster, we take as a quantitative criterium that they should have a crossing time several times smaller than the age of the youngest embedded clusters. With a typical velocity dispersion of 0.9 km.s$^{-1}$, and a crossing time of 1/3 of 1~Myr, the obtained distance is $\sim 0.3\,$pc. The observed MDCs of Cygnus X are all embedded in larger scale clumps which have total masses ranging from 260 to 7300$\,$\msol$\,$, sizes ranging from 0.25 to 0.74$\,$pc (FWHM), and average densities ranging from $2.5 \times 10^4$ to $1.3 \times 10^5\,$cm$^{-3}$ (see Tab.~2 in \citealp{motte2007}). It is therefore of the order of several hundreds of solar masses of gas at relatively high density which is available in a radius of $\sim 0.3\,$pc around each MDC which should be enough to form a large population of low-mass stars in their surroundings. 
Alternatively, a large, already formed population of low-mass stars and protostars might be present in the immediate surroundings of the MDCs. The low-mass stars and protostars are difficult to observe at the distance of Cygnus X and it is therefore difficult to establish if there is already such a large enough population of low-mass counterparts in the surroundings. We will investigate this second possibility in a more complete way in a forthcoming paper using the recent Spitzer data from the legacy program towards Cygnus X \citep{hora2009}.

\section{Conclusions}

We have been imaging at high spatial resolution with the IRAM PdBI interferometer the most massive IR-quiet dense cores discovered by \citet{motte2007} in Cygnus X. These MDCs are among the best targets to investigate the origin, and the earliest phases of the formation of high-mass stars and of OB clusters. From this unique dataset, we could reach the following conclusions and results: 
   \begin{enumerate}
      \item All except one MDC (CygX-N40) were found to be bright and easily detected with the PdBI in a	beam of $\sim 3.0$\asec $\times$~2.3\asec at 3.5~mm, and $\sim 1.1$\asec $\times$~0.9\asec at 1.3~mm.
      \item There is a striking morphological difference between the lower spatial resolution maps at 3.5~mm and the high resolution maps at 1.3~mm for 3 out of the 5 MDCs. While at 3.5~mm, the MDCs are dominated by bright central emissions, at 1.3~mm, these emissions clearly split into a population of compact fragments with typical separations of 5000~AU and sizes larger than 1000~AU.
      \item All except  one MDC (CygX-N63) are sub-fragmented. CygX-N63 consists of a single, centrally concentrated object down to the smallest scales. It could actually be a single massive protostar with an envelope mass as large as $\sim60\,$\msol$\,$ and may therefore form an O star.
      \item A population of 23 compact dense fragments are recognized as certainly self-gravitating, proto-stellar objects. Among them, 9 objects are good candidates to be true, individual high-mass protostars in the sense that they are precursors of OB stars.
      \item The targeted MDCs of Cygnus X have globally the properties proposed as initial conditions by \citet{mckee&tan2003} for the turbulence-regulated, monolithic collapse. But in contrast to this scenario, these MDCs are mostly sub-fragmented. Only CygX-N63 is found to be single. To further investigate whether this particular core could have evolved towards global collapse by a slow evolution, regulated by a true turbulent field, a detailed study at high spatial resolution is mandatory to, for instance, probe its kinematical structure, and the nature of its velocity dispersion.
      \item On the other hand, the properties of the fragmentation of the MDCs are not in agreement with a pure gravo-turbulent fragmentation.  A large fraction of the mass of 3 out 5 MDCs is located in a few, high-mass objects which would not be expected in the Jeans mass regulated fragmentation. The local Jeans masses are too low, of the order of 1$\,$\msol$\,$. Similarly, the core mass function of the fragments is globally (for the 5 MDCs) biased towards high-masses as compared with a normal IMF-like core mass function.
      \item We found a clear trend that the fragmentation efficiency, defined as the total fraction of mass in self-gravitating fragments, is increasing with the average density of the cores. Since the density is directly related to self-gravity, we interpret this result as due to the larger effect of self-gravity in the densest cores leading to global collapse of the most extreme cases.
      \item In the same time, as the density increases, the fragmentation tends to lead to a fewer number of predominantly high-mass fragments in the central part of the cores. The central parts of at least 3 ot the 5 MDCs are so clearly dominated by massive objects that they are certainly leading to regions with a clear mass segregation. This mass segregation is proposed to correspond to the expected primordial mass segregation of young stellar clusters.      
   \end{enumerate}

\begin{acknowledgements}
      Part of this work was supported by the French 
      \emph{Agence Nationale pour la Recherche, ANR\/} project 
      "PROBeS", number ANR-08-BLAN-0241.
\end{acknowledgements}

\bibliography{biblio}
\bibliographystyle{aa}

\end{document}